\documentclass[iop]{emulateapj}
\usepackage{latexsym}
\usepackage{dcolumn}% Align table columns on decimal point
\usepackage{bm}% bold math
\usepackage{amsmath}
\usepackage{natbib}
\input{epsf}
\newcommand{\be}{\begin{equation}}
\newcommand{\ee}{\end{equation}}
\newcommand{\ba}{\begin{eqnarray}}
\newcommand{\ea}{\end{eqnarray}}
\newcommand{\bi}{\begin{itemize}}
\newcommand{\ei}{\end{itemize}}
\newcommand{\bfi}{\begin{figure}[t]
\epsfxsize=9cm
\epsffile}
\newcommand{\bfig}{\begin{figure*}[t]
\epsfxsize=15cm
\epsffile}
\newcommand{\efi}{\end{figure}}
\newcommand{\efig}{\end{figure*}}
\newcommand{\no}{\nonumber}

\newcommand{\mpch}{{\rm Mpc}/h}

\begin{document}
\title{Weak lensing power spectrum reconstruction by counting
  galaxies.-- I: the ABS method}
\author{Xinjuan Yang$^1$, Jun Zhang$^2$, Yu Yu$^3$, Pengjie Zhang$^{2,4}$\\
$^1$ Institute for Advanced Physics \& Mathematics, Zhejiang
University of Technology, Hangzhou, 310032, China\\
$^2$ Department of astronomy, Shanghai Jiao Tong
  University, 955 Jianchuan road, Shanghai, 200240\\
$^3$ Key Laboratory for Research in Galaxies and Cosmology, Shanghai
  Astronomical Observatory, 80 Nandan Road, Shanghai, 200030,
  China\\
$^4$ IFSA Collaborative Innovation Center, Shanghai Jiao Tong
University, Shanghai 200240, China
}
\email[Email me at: ]{zhangpj@sjtu.edu.cn}
\begin{abstract}
We propose an Analytical method of Blind Separation (ABS) of cosmic
magnification from the intrinsic fluctuations of galaxy number density in the observed galaxy
number density distribution. The ABS method utilizes the different
dependences of the signal (cosmic magnification) and contamination
(galaxy intrinsic clustering) on galaxy flux, to separate the two. It
works directly on the measured cross galaxy angular power spectra
between different flux bins. It determines/reconstructs the lensing power 
spectrum analytically, without assumptions of galaxy intrinsic
clustering and cosmology. It is unbiased in the limit of infinite number of
galaxies. In reality the lensing reconstruction accuracy 
depends on survey configurations, galaxy
biases, and other complexities, due to finite number of galaxies and the
resulting shot noise fluctuations in the cross galaxy power spectra.  We estimate
its performance (systematic and 
statistical errors) in various cases. We find that, stage IV dark
energy surveys such as SKA and LSST are capable of reconstructing the lensing 
power spectrum at $z\simeq 1$ and $\ell\la 5000$ accurately. This lensing reconstruction only requires counting galaxies, and is therefore highly complementary to the cosmic 
shear measurement by the same surveys. 
\end{abstract}
\keywords{cosmology: observations: large-scale structure of universe: dark
  matter: dark energy}
\maketitle

\section{introduction}
Weak gravitational lensing probes the large scale
structure and geometry of the universe
\citep{2001PhR...340..291B,2003ARA&A..41..645R,Hoekstra08,Munshi08}. It
contains valuable information of fundamental physics, such as dark matter, dark energy and  the nature of
gravity at cosmological scales  
\citep{2001PhR...340..291B,2003ARA&A..41..645R,DETF,Hoekstra08,Munshi08,2013PhR...530...87W}. A
major goal of precision cosmology is to both measure and model weak lensing
accurately. 

Despite of being a weak cosmological signal,  weak lensing can be
measured and has been measured in various ways. (1) The most
comprehensively studied method  is to measure the  lensing distorted
galaxy shapes. Existing surveys such as CFHTLenS
~\citep{2012MNRAS.427..146H,Kilbinger:2012qz,2014MNRAS.441.2725F},
SDSS ~\citep{2014MNRAS.440.1322H} and RCSLenS
\citep{2016MNRAS.463..635H} have
already measured this cosmic shear signal robustly. The measurement
precision will be further improved by ongoing surveys
such as KiDS\footnote{http://kids.strw.leidenuniv.nl}, DES\footnote{https://www.darkenergysurvey.org/} and
HSC\footnote{http://www.naoj.org/Projects/HSC/index.html}. For
example, KiDS has presented intermediate
results based on observations of 450 deg$^2$ \citep{2017MNRAS.465.1454H}.  DES
has also  measured cosmic shear in its science
verification data
\citep{2015PhRvL.115e1301C,2016MNRAS.460.2245J,2016PhRvD..94b2002B}. Eventually, the planned wider/deeper stage IV surveys such as
Euclid\footnote{http://sci.esa.int/euclid/} and
LSST\footnote{http://www.lsst.org/lsst/} will achieve $1\%$ or better precision. (2) Weak lensing distorts
CMB sky \citep{Seljak96}, which enables the reconstruction of lensing
deflection over the sky
\citep{1999PhRvL..82.2636S,2003PhRvD..67h3002O}. Detection of this CMB
lensing effect in the WMAP temperature-galaxy  cross correlations was
first reported by \citet{2007PhRvD..76d3510S,2008PhRvD..78d3520H}.
Cross correlations between lensing B-mode polarization and the large scale structure
have been detected recently
(SPTpol, \citet{2013PhRvL.111n1301H}; POLARBEAR,
\citet{2014PhRvL.112m1302A,2014PhRvL.113b1301A}; ACTpol,
\citet{2015ApJ...808....7V}). The lensing auto power spectrum, which is
cosmologically more useful, has been reconstructed by ACT
\citep{2011PhRvL.107b1301D}, SPT \citep{2012ApJ...756..142V}, POLARBEAR
\citep{2014PhRvL.113b1301A}, and Planck
\citep{2014A&A...571A..17P,2016A&A...594A..15P}.  In particular Planck
has achieved the most significant detection at a level of
$40\sigma$. (3) The lensing induced magnification in galaxy/supernovae flux and
size provides an alternative to cosmic shear
(e.g. \citet{2006PhRvD..74f3515D,Cooray06a,Vallinotto11,2013MNRAS.433L...6H,2014ApJ...780L..16H,2014MNRAS.437.2471D,2015MNRAS.452.1202A,2015ApJ...806...45Z}). It
has enabled recent detections (e.g. 
\citet{2012ApJ...744L..22S,2016MNRAS.457..764D}). 
%It also invokes studies
%of possible systematics such as selection effects and intrinsic
%size spatial correlation
%\citep{2013MNRAS.430.2844C,2015MNRAS.449.2059C,2015MNRAS.454..478J}. 

Gravitational lensing not only distorts galaxy images, but also
changes the spatial distribution of galaxies
\citep{1995A&A...298..661B,2001PhR...340..291B}. In principle, this cosmic
magnification (magnification bias) effect can also be used to measure
weak lensing. It has several appealing advantages. It is free of point
spread function (PSF) and intrinsic alignment that cosmic shear
measurement suffers severely
(\citet{2015MNRAS.450.2963M,2015PhR...558....1T} and references
therein). It has less stringent requirements on galaxy
observation. For example, cosmic shear measurement require galaxies to
be sufficiently large and bright. But in principle smaller or fainter
galaxies, as long as above 
the detection limit, can all be used for  cosmic magnification measurement
\citep{Zhang05,Zhang06a}. Therefore for the same survey there will be significantly more galaxies useful for cosmic magnification measurement than for
cosmic shear measurement. 

Nevertheless,  in reality cosmic magnification (magnification bias) is
usually overwhelmed by intrinsic fluctuations in galaxy number
density. And even worse, such intrinsic galaxy density fluctuations
are spatially correlated,  even over scales of $\sim 100\mpch$. This
galaxy intrinsic clustering is the major obstacle of lensing
reconstruction through cosmic magnification.  It is
analogous to the intrinsic alignment in cosmic shear measurement, but
more severe. To avoid
this problem, cosmic magnification is usually measured indirectly, through 
cross correlations between background populations such as quasars and
foreground galaxies. It was first detected using SDSS galaxies as
foreground population and SDSS quasars as background population
\citep{2005ApJ...633..589S}. It was then detected using various other data
sets
\citep{2009A&A...507..683H,2010MNRAS.405.1025M,2011MNRAS.414..596W,2012MNRAS.426.2489M,2013MNRAS.429.3230H,2014MNRAS.440.3701B,2014MNRAS.442.2680G,2016arXiv161110326G}. It
can also be detected in galaxies behind galaxy clusters (e.g.
\citet{2014MNRAS.439.3755F,2016MNRAS.457.3050C}) and  in  high redshift galaxies
(e.g. \cite{2011Natur.469..181W}), where the lensing
magnification is stronger. 

However, since the above cross correlation measurements are subject to the (foreground) galaxy bias,  their cosmological applications are
limited. For the purpose of precision cosmology, the measurement of the
lensing auto correlation instead of galaxy-galaxy lensing is more
desired. To do so, we need to separate the magnification bias from
the galaxy intrinsic clustering, both at map level or at the level of
two-point correlation of the same redshift bin. \citet{Zhang05} first pointed out
that this is in principle feasible, since magnification bias and the intrinsic galaxy
clustering depend differently on galaxy flux. The intrinsic
  clustering is proportional to the galaxy bias, in which the dominant
  part (the deterministic bias) is always positive. In contrast,
  the magnification  bias changes sign from faint end to bright
  end. Therefore it is possible to
  separte the two in flux space. \citet{YangXJ11} implemented this idea with an iterative
solver to reconstruct the lensing convergence map from the surface
number density distribution of galaxies within a redshift
bin. \citet{YangXJ15} implemented the same idea, but with the aim of reconstructing
the lensing power spectrum, which is the most important lensing
statistics. The major finding is a mathematical proof  that the
lensing power spectrum can be uniquely determined under the condition of
deterministic bias. This lensing reconstruction requires no  further
assumptions on the galaxy bias.  

A long standing unresolved problem in lensing reconstruction through
cosmic magnification is the  stochasticity in the galaxy
intrinsic clustering. \citet{YangXJ11,YangXJ15} found that it is the most significant
limiting factor.  One possible solution is to model it and then
marginalize it in parameter fitting. However, being connected with the
nonlinear structure formation  and complicated process of galaxy
formation, theoretical understanding  of the galaxy stochasticity is
highly challenging and uncertain. Ideally we shall reconstruct weak lensing without priors or
modelling of galaxy intrinsic clustering, including its stochasticity. 

This paper is the first in a series of papers presenting the ABS method to solve this crucial
problem. It no longer requires vanishing stochasticity. It works for
general case of galaxy intrinsic clustering.  {\bf ABS} in this paper
stands for {\bf A}nalytical method of {\bf B}lind
{\bf S}eparation of cosmic magnification from the galaxy intrinsic
clustering. It is
blind in that it relies on no assumptions on the galaxy intrinsic
clustering. It is analytical in that the lensing power spectrum is
determined by an analytical formula that we recently discovered,
instead of numerical fittings of many unknown parameters.  It is
unbiased when the survey is sufficiently powerful that shot noise in
the observed galaxy distribution can be well controlled. Here we  advertise that
the ABS method has other  applications. We have demonstrated its power
in CMB B-mode foreground 
removal \citep{ABS}.

This paper is organized as follows. We describe the methodology in \S
\ref{sec:ABS}. We demonstrate that its applicability is within the reach
of stage IV dark energy projects such as LSST and SKA (\S \ref{sec:test}). We carry out
much more comprehensive tests to further
demonstrate its generality  in \S \ref{sec:test2}.  We
discuss and summarize in \S \ref{sec:summary}. We caution that many
tests are against highly hypothetical cases, and are only 
served for evaluating the generality of the ABS method. In a companion
paper we will combine with N-body simulations and galaxy mocks to more reliably quantify its
realistic  performance for realistic surveys.

\section{The ABS method}
\label{sec:ABS}
 In the weak lensing
regime, the galaxy number overdensity after lensing is given by (e.g. \citet{1995A&A...298..661B})
\be
\label{eqn:magnification}
\delta^L_g=g \kappa+\delta_g\ .
\ee
Here $\kappa$ is the lensing convergence that we want to measure, and $\delta_g$ is the
intrinsic galaxy number overdensity that we want to eliminate. The
prefactor $g=2(\alpha-1)$ is determined by
$n(F)$, the average number of galaxies per flux interval. For a narrow flux bin, $\alpha\equiv -d\ln n/d\ln
F-1$.\footnote{Notice that this is slightly different to the
  originally derived  
  expression for the number density fluctuation over a flux threshold
  \citep{1995A&A...298..661B}. In that case, 
  $\alpha=-d\ln N/d\ln F$, with $N(F)=\int_F^{\infty} n(F)dF$ as the number of galaxies
  brighter than $F$. }  Cosmic magnification (magnification bias) has a specific
flux dependence $g(F)$, which is in general different to the flux
dependence of the contamination ($\delta_g$). Therefore in principle
the signal and the contaminations can be
separated in flux space. This motives us to split galaxies in a
redshift bin into $N_F$ flux/luminosity
bins and then carry out weak lensing reconstruction in flux space
\citep{Zhang05,YangXJ11,YangXJ15}.  We denote $g_i$
($i=1,\cdots, N_F$) as the 
corresponding $g$ of the $i$-th flux bin.  We work in the Fourier
space and focus on the determination of lensing power spectrum
$C_\kappa(\ell)$ at multipole $\ell$. 

\subsection{The case of ideal measurement}
First we consider ideal measurement, in which measurement
errors in galaxy clustering (e.g. shot noise) are negligible. 
For each multipole $\ell$ bin, we will have 
$N_F^2$ angular power spectra $C^L_{ij}$ between the $i$-th and $j$-th flux bins, 
\ba
\label{eqn:CL}
C^L_{ij}(\ell)&=&g_ig_jC_\kappa (\ell)\no\\
&&+(b^{\rm D}_i(\ell) g_j+b^{\rm D}_j (\ell) g_i)C_{m\kappa}(\ell)+C^g_{ij}(\ell)\ .
\ea 
These cross power spectra form a symmetric matrix of order $N_F$, with
$N_F(N_F+1)/2$ independent components.   
$C^{g}_{ij}(\ell)$ is the power spectrum of galaxy intrinsic
clustering. $b^{\rm D}_i(\ell)$ is the deterministic bias of galaxies in the
$i$-th flux bin. Notice that it has the superscript ``D'' and should
not be confused with the linear and quadratic bias $b^{(1,2)}$ introduced later in the paper.   We explicitly show the scale ($\ell$) dependence of bias and
emphasize that we do not make any assumptions on this scale
dependence. Since the source distribution of galaxies is not
infinitesimally thin, it overlaps with the lens distribution. Therefore the intrinsic number
overdensity $\delta_g$
of galaxies in the $i$-the ($j$-th) flux bin is spatially correlated
with the magnification bias of galaxies in the $j$-th ($i$-th) flux
bin. The stochastic part of $\delta_g$ is uncorrelated to the matter
overdensity $\delta_m$ and is therefore uncorrelated to the lensing
convergence $\kappa$.  Therefore, the cross correlation is
proportional to the deterministic galaxy bias $b^{\rm D}(\ell)$. It is also
proportional to the cross power spectrum $C_{m\kappa}$ between the
matter distribution and the lensing 
convergence. 

$C_\kappa$ in the first line of Eq. \ref{eqn:CL} is the signal that we
want to measure/reconstruct, and all terms in the second line are contaminations
that we want to  eliminate.  However, due to an intrinsic degeneracy found in \cite{YangXJ15}, no unique
solution of $C_\kappa$ exists.  Fortunately, \citet{YangXJ15} found
that  Eq. \ref{eqn:CL} can be rewritten as
\be
\label{eqn:Cij1}
C^L_{ij}=g_ig_j\tilde{C}_\kappa+\tilde{C}^g_{ij}\ ,
\ee
with
\ba
\label{eqn:transformation}
\tilde{C}_\kappa &\equiv& C_\kappa(1-r^2_{m\kappa})\ ,\
r^2_{m\kappa}\equiv \frac{C^2_{m\kappa}}{C_mC_\kappa}\ ,\no\\
\tilde{C}^g_{ij}&\equiv&
C^g_{ij}+\tilde{b}_{i}\tilde{b}_{j}-b^{\rm D}_ib^{\rm D}_jC_m\ , \no
\\
\tilde{b}_i&\equiv& \sqrt{C_m}(b^{\rm D}_i+g_i \frac{C_{m\kappa}}{C_m})\ .
\ea
\citet{YangXJ15} proved that, under the condition of deterministic
bias, the solution to $\tilde{C}_\kappa$ is unique. In the current
paper, we will show that the uniqueness of solution holds even with
the existence of stochastic bias. Even better, there exists an
analytical solution of $\tilde{C}_\kappa$. 

The intrinsic clustering matrix $C^g_{ij}$ has order $N_F$, but its
rank $M$ (the number of eigenmodes) depends on the degrees of freedom in the intrinsic
clustering. For example, if the intrinsic clustering is fully
deterministic, its rank $M=1$. In reality, galaxy clustering contains
stochasticities, so $M>1$. Numerical simulations suggest that $C^g_{ij}$ has
only limited degrees of freedom.  We only need $2$ or $3$
eigenmodes to describe the matrix $C^{\rm halo}_{ij}$, the
cross power spectra of halos between different halo mass bins
\citep{Bonoli09,Hamaus10}. Namely for halos, $M=2$ or $3$. For
galaxies, there is so far no quantitative investigation. However, from
the viewpoint of halo model, we expect no fundamental difference
between $C^g_{ij}$ and $C^{\rm halo}_{ij}$.  The deterministic bias
vector ${\bf b}^{\rm D}$ is a linear combination of eigenvectors of
${\bf C}^g$. Then by linear algebra, the rank
of matrix $C^L_{ij}$ is $M+1$, due to the extra linearly independent
vector ${\bf g}$ in Eq. \ref{eqn:CL}. To identify all these eigenmodes, we
need the number of flux bin $N_F\geq M+1$.  

The mathematical structure of Eq. \ref{eqn:Cij1} is identical
to Eq. 1 in a recent paper by two of the authors \citep{ABS}. This
paper proposes an Analytical method of Blind Separation (ABS) of CMB
B-mode from foregrounds. There ABS works on the cross band powers
between CMB frequency bands, 
\be
\label{eqn:CMB}
\mathcal{D}_{ij}=f_i^{\rm B}f_j^{\rm B}\mathcal{D}_{\rm
  B}+\mathcal{D}^{\rm fore}_{ij} \ .
\ee
It solves for the B-mode
band power $\mathcal{D}_{\rm B}$, without assumptions on the
foreground band power $\mathcal{D}^{\rm fore}_{ij}$. ABS is made
possible by two facts. One is that the CMB has a known blackbody
frequency dependence $f^{\rm B}_i\equiv f^{\rm B}(\nu_i)$. The other is that in principle we
can have more frequency bands than the number of independent
foreground components. By the correspondences of
\be
(F, C^L_{ij},g_i,\tilde{C}_\kappa,\tilde{C}^g_{ij})\longleftrightarrow
(\nu, \mathcal{D}_{ij}, f_i^{\rm B}, \mathcal{D}_{\rm B}, \mathcal{D}^{\rm
  fore}_{ij})\ ,
\ee 
Eq. 3 and Eq. \ref{eqn:CMB} (namely Eq. 1 in \citet{ABS}) are indeed mathematically
identical. Furthermore, as $f_i^{\rm B}$ is observationally
known in the case of CMB, $g_i$ is observationally known in the case of cosmic
magnification. Therefore, the ABS method applies
to both cases. So we will simply re-express the results in \citet{ABS}
in the language of weak lensing. 
\bi
\item {\bf The solution to $\tilde{C}_\kappa$
is unique, as long as $M+1\leq N_F$}. Then by studies of halo clustering
using numerical simulations, once we choose $N_F\geq 4$, weak lensing
power spectrum ($\tilde{C}_\kappa$) reconstruction is expected to be
unique. 
\item  {\bf There exists the following analytical solution for $\tilde{C}_\kappa$},
\be 
\label{eqn:analyticalsolution}
\tilde{C}_{\kappa}=\frac{1}{G^TE^{-1}G}=\left(\sum_{\mu=1}^{M+1} G_\mu^2\lambda_\mu^{-1}\right)^{-1}\ . 
\ee 
Here, $\lambda_\mu$ ($\mu=1,\cdots, M+1$) is the eigenvalue of the
$\mu$-th eigenvector ${\bf E}^{(\mu)}$. $E_{\mu\nu}\equiv {\bf
  E}^{(\mu),T}\cdot {\bf C}^L\cdot {\bf
  E}^{(\nu)}$ is the projection of the matrix ${\bf C}^L$ onto the
$M+1$ dimension  space of eigenvectors.  $G_\mu\equiv
{\bf g}\cdot{\bf E}^{(\mu)}$. The last expression requires ${\bf
  E}^{(\mu)}\cdot{\bf E}^{(\mu)}=1$ and we will adopt this
normalization throughout the paper. 
\ei

There is one technical issue on the
$C_\kappa$-$\tilde{C}_\kappa$ relation to clarify here. The relative
difference between the two is $r^2_{m\kappa}$, where $r_{m\kappa}$ is 
the cross correlation coefficient between the distribution of source
and the lensing field. Usually $r^2_{m\kappa}\ll 1$ \citep{YangXJ15}. For example, for $1.0<z<1.2$, $r^2_{m\kappa}\la 
2\times 10^{-3}$. Therefore $\tilde{C}_\kappa=C_\kappa$ is an
excellent approximation for narrow redshift distribution. However,
when the width of redshift distribution increases, $r^2_{m\kappa}$
increases. For example, for $0.8<z<1.2$, 
$r^2_{m\kappa}\simeq 0.01$ over a wide range of $\ell$.  In this case, we may no longer take
$\tilde{C}_\kappa=C_\kappa$, and have to keep in mind that the
reconstructed one is
$\tilde{C}_\kappa=C_\kappa(1-r^2_{m\kappa})$. Fortunately, this does
not introduce any new uncertainty in cosmological constraints. The
reason is that $r_{m\kappa}$ does not 
depend on the galaxy bias and can be calculated without
introducing extra uncertainties than in $C_\kappa$ calculation. This
means that  $\tilde{C}_\kappa$ is 
essentially identical to $C_\kappa$ in cosmological applications.

\subsection{Including measurement errors}
In reality, the measured (lensed) galaxy clustering is contaminated by
shot noise, due to finite number of galaxies. The ensemble
average of the shot noise power spectrum can be predicted and
subtracted from the observation. What left is 
\be
\label{eqn:CLobs}
C^{\rm obs}_{ij}=C^L_{ij}+\delta C_{ij}^{\rm shot}\ .
\ee
$\delta
C_{ij}^{\rm shot}$ is the residual shot noise due to statistical fluctuations. It has the following properties,
\ba
\label{eqn:shotnoise}
\langle \delta C_{ij}^{\rm shot}\rangle&=&0 \ ,\no\\
\langle \delta C^{\rm
  shot}_{ij}\delta C^{\rm
  shot}_{km}\rangle&=&\frac{1}{2}\sigma_i\sigma_j(\delta_{ik}\delta_{jm}+\delta_{im}\delta_{jk})\
.
\ea
Here,
$\sigma_i=(4\pi f_{\rm sky}/N_i)\times\sqrt{2/[(2\ell+1)\Delta \ell
  f_{\rm sky}]}$ is the statistical error caused by shot noise in the
band power in the multipole range of $\ell-\Delta \ell/2$ and $\ell+\Delta \ell/2$. $f_{\rm sky}$ is the fractional
sky coverage and $N_i$ is the total number of galaxies in the $i$-th
flux bin. Hereafter we will choose the flux bin sizes such that
$N_1=N_2=\cdots=N_{N_F}=N_{\rm tot}/N_F$, where $N_{\rm tot}$ is the
total number of galaxies in all flux bins. We then have
$\sigma_1=\sigma_2=\cdots$, which we denote as $\sigma_{\rm shot}$.
\be
\label{eqn:sigmashot}
\sigma_{\rm shot}=\left(\frac{4\pi f_{\rm sky}}{N_{\rm tot}/N_F}\right)\times\sqrt{\frac{2}{(2\ell+1)\Delta \ell
  f_{\rm sky}}}\ .
\ee

With the presence of shot noise, the rank of matrix
$C_{ij}^{\rm obs}$ will be equal to its order $N_F$. Surprisingly,
Eq. \ref{eqn:analyticalsolution} can still be implemented in the data
analysis, as shown for the case of CMB B-mode \citep{ABS}.  Only one
straightforward modification is needed to  account for shot noise. 
\bi
\item {\bf Step 1}. We compute all $N_F$ eigenmodes of $C^{\rm obs}_{ij}$. 
\item  {\bf Step 2}. We measure $\tilde{C}_{\kappa}$ from 
  Eq. \ref{eqn:analyticalsolution}, {\it but  only using eigenmodes with
$\lambda_\mu>\lambda_{\rm cut}$}.  
\ei
Here, $\lambda_{\rm cut}$ is a cut adopted to filter away unphysical
eigenmodes caused by statistical fluctuations of shot noise.  Residual
shot noise (and other measurement statistical error in general) not only affects the determination of
physical eigenmodes, but also 
induces unphysical eigenmodes with eigenvalues of typical amplitude
$\sigma_{\rm shot}$. This roughly sets the value of $\lambda_{\rm
  cut}\sim \sigma_{\rm shot}$. Nevertheless, the above recipe may
still miss physical eigenmodes, or fail to exclude unphysical eigenmodes. The
impacts will be better understood in the language of
$\lambda_\mu$-$c_\mu$ diagnostic of \S 
\ref{subsec:lambdac}.  The
resulting systematic error will be 
quantified numerically through our simulated data (\S \ref{sec:test}
\& \ref{sec:test2}). The ambiguity in the choice of $\lambda_{\rm
  cut}$ will be discussed in \S
\ref{subsec:cut}.

The above method  of measuring $\tilde{C}_{\kappa}$, even including
the determination of $M$,  is completely fixed by the data, and relies
on no priors of galaxy intrinsic clustering. Furthermore, it has a precious property
that it is unbiased, in the limit of low measurement errors.

\subsection{The $\lambda_\mu$-$c_\mu$ diagnostic}
\label{subsec:lambdac}
The ABS
method automatically utilizes the unique (and known) flux dependence
of the lensing signal to blindly separate it
from overwhelming contaminations of galaxy intrinsic clustering.
According to Eq. \ref{eqn:analyticalsolution}, different eigenmode of the
matrix $C^L_{ij}$ has different contribution to the lensing
reconstruction. The contribution of the $\mu$-th physical eigenmode is 
\be
c_\mu\equiv \frac{G_\mu^2/\lambda_\mu}{\sum_{\alpha=1}^{M+1}
  G_\alpha^2/\lambda_\alpha}\ .
\ee
The value of $\lambda_\mu$ determines whether a survey can detect this
eigenmode, and the value of $c_\mu$ determines whether this eigenmode
is relevant for the lensing reconstruction. Statistical error (shot
noise as we consider here) in $C^{\rm obs}_{ij}$ may prohibit
identification of physical eigenmode, and generate unphysical
eigenmodes. The former leads to overestimation of $\tilde{C}_\kappa$,
while the later leads to underestimation.  This $\lambda_\mu$-$c_\mu$
diagnostic is developed in \citet{ABS}.

When the measurement error ($\sigma_{\rm shot}$) is small,  all physical eigenmodes can be
identified. But there exists the possibility of unphysical  eigenmodes
with eigenvaues exceeding $\lambda_{\rm cut}$. When a unphysical
eigenmode is wrongly included,  it results in underestimation by a factor
$-c_\mu/(1+c_\mu)$. 

In contrast, when $\sigma_{\rm shot}$ is large and we may fail to
detect some physial eigenmodes with small eigenvalues.  Missing the $\mu$-th component
causes overestimation by a factor of $c_\mu/(1-c_\mu)$.  In reality, only those eigenmodes with significant
$c_\mu$ (e.g. $c_\mu>0.01$) are relevant for lensing determination. In
another word, to achieve $1\%$ level accuracy in the reconstruction, all
eigenmodes with $c_\mu>0.01$ have to be detected. If one of these
eigenmodes has  $\lambda<\lambda_{\rm cut}\sim \sigma_{\rm shot}$, it will be missed in the
reconstruction, resulting in significant bias in the reconstructed
$\tilde{C}_\kappa$.

\subsection{Statistical error}
Following similar derivation in \citet{ABS} and in the limit of small
shot noise fluctuations, we derive the r.m.s. error  induced by shot
noise, 
\ba
\label{eqn:error}
\sigma_{C_\kappa}&\equiv& \langle \delta
  \tilde{C}_\kappa^2\rangle^{1/2}=\eta\times \sigma_{\rm shot} \ ,\no\\
\eta &\equiv & \left(\sum_{\mu=1}^{M+1}\frac{G_\mu^2}{\lambda_\mu^2}\times
    \tilde{C}_\kappa^2\right)
  \ .
\ea
As expected, the statistical error $\sigma_{C_\kappa}\propto
\sigma_{\rm shot}$. The prefactor  $\eta$ is dimensionless,  determined by the interplay
between the cosmic magnification and the intrinsic clustering,
implicitly through $\lambda_\mu$ and $G_\mu\equiv {\bf 
  g}\cdot {\bf E}^{(\mu)}$. Eq. \ref{eqn:error} quantifies the
possibility of measuring the lensing power spectrum in the given sky
coverage and source redshift, by counting
galaxies.\footnote{Eq. \ref{eqn:error}  does not include
cosmic variance of the lensing field. This source of statistical error
is needed and only needed when we compare the measured lensing power
spectrum with the theoretically predicted ensemble average to obtain
cosmological parameter constraints.}

One immediate conclusion from Eq. \ref{eqn:analyticalsolution} \&
\ref{eqn:error} is that eigenvectors orthogonal to ${\bf g}$ not only  have no
contribution to the weak lensing reconstruction, but also have no
impact on the statistical error of  reconstruction. The best case that
we can expect for the lensing reconstruction  is that all other
eigenvectors are orthogonal to ${\bf g}$. In this case, one
eigenvector is $\hat{g}\equiv {\bf
  g}/g$, where $g\equiv \sqrt{\sum_i g_i^2}$. The corresponding
eigenvalue is $\sum_i
g_i^2\tilde{C}_\kappa$. All other eigenmodes 
have $G=0$. We then have $\eta=1/g^2$. This sets up the lower limit of
statistical error of the weak lensing reconstruction, $\sigma_{\rm
  shot}/g^2$. In general there will be other eigenvectors unorthogonal to
${\bf g}$, so the actual statistical error is larger than the above
value.

\section{Testing the ABS method against the fiducial case}
\label{sec:test}
The ABS method is unbiased in reconstructing the lensing power
spectrum, under the condition of vanishing measurement error in the
observed galaxy clustering (e.g. shot noise). In reality, this
condition is violated due to limited number of  galaxies (and
possibly other observation complexities). This causes the realistic performance of the ABS
method to depend on many factors, such as survey specifications which
determine $\sigma_{\rm shot}$, galaxy biases, redshift range, the
number of flux bins and cosmology. 

We expect $\sigma_{\rm shot}$ and the galaxy biases as the most
important factors affecting the ABS performance. Therefore, for brevity
we will fix the cosmology, the number of flux bins, and the redshift
range throughout the paper. (1)  We adopt a flat $\Lambda$CDM cosmology with $\Omega_m=0.26$， 
$\Omega_{\Lambda}=1-\Omega_m$, $\Omega_b=0.044$, $h=0.72$, $n_s=0.96$ and
$\sigma_8=0.8$. The weak lensing angular (2D) power spectrum is calculated by the
Limber integral, in which the 3D nonlinear matter power spectrum is
calculated using the halofit fitting formular \citep{Smith03}.  (2)
 We fix the redshift bin $0.8<z<1.2$. This is the
source redshift range accessible by many cosmic shear surveys such as
DES, HSC and LSST. It is therefore convenient for comparison between
cosmic shear and cosmic 
magnification.  (3) We fix the number of flux
bins as $N_F=5$. This is implied (but not fixed) by the following considerations.  First, the
ABS method requires $N_F\geq M+1$. For most cases that we test,
$M=2$. Therefore $N_F=3$ is the minimal requirement. Second,  we
prefer a larger $N_F$ (finer flux bin size) to better capture the
different flux dependences of cosmic magnification and galaxy bias.
However,  $N_F$ can not be arbitrarily large because of increasing
shot noise per flux bin with increasing $N_F$.  We then take $N_F=5$ as our
first try.  A remaing question of importance is the optimal choice of
$N_F$.  In future we will test
the ABS method using mocks generated from N-body simulations. At that
stage, we will carry out more thorough tests of the ABS method against
other redshift range, other cosmology, and figure out the optimal
choice of $N_F$.

\begin{figure}[t]
\epsfxsize=9cm
\epsffile{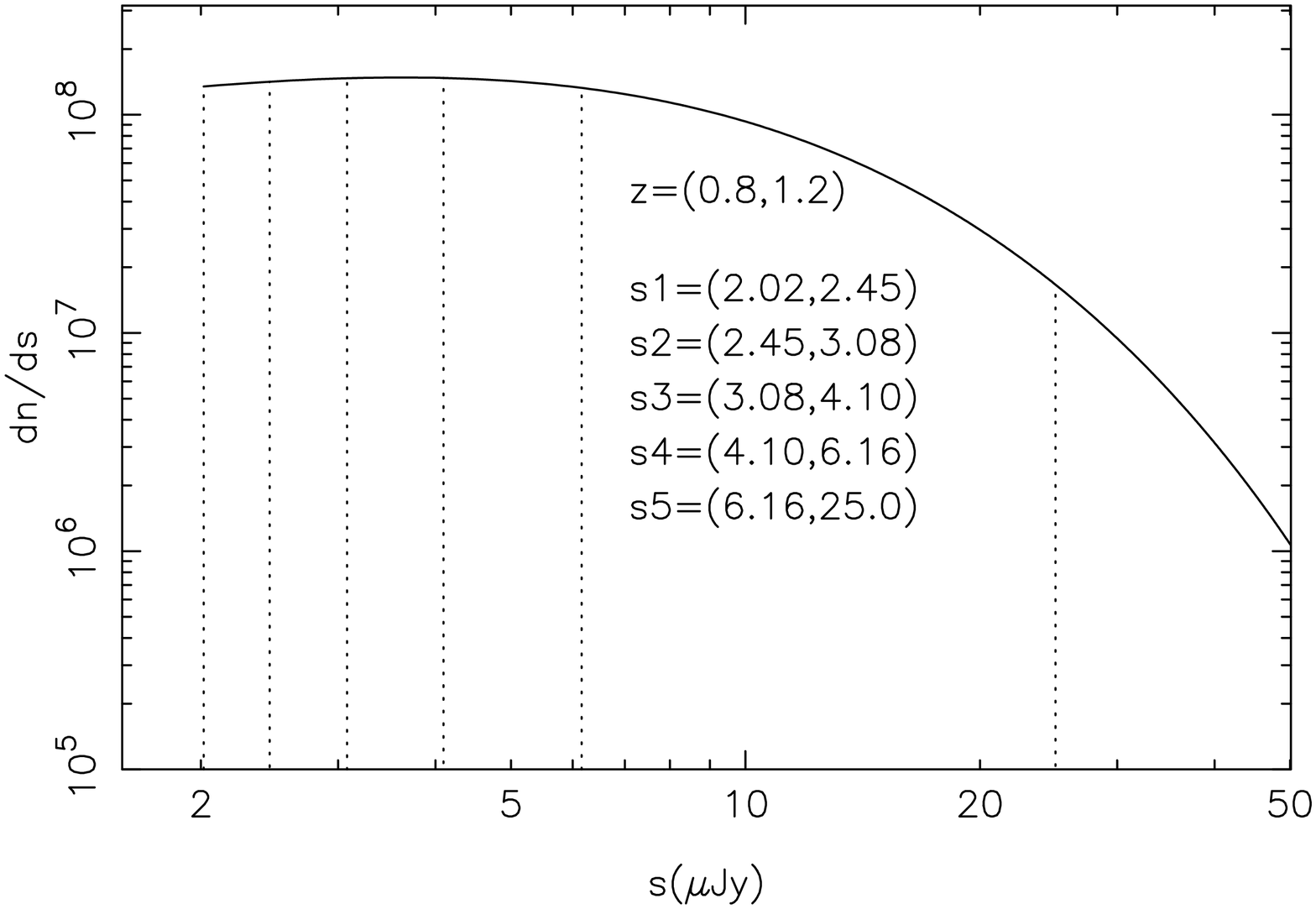}
\caption{The fiducial number distribution of galaxies (solid line) at $0.8<z<1.2$ as
  a function of 21cm flux $s$ ($\mu$Jy) in the fiducial SKA-like
  survey. The vertical dot lines show the ranges 
  of the adopted $5$ flux bins. More explanations are provided in \S
  \ref{sub:fiducialsurvey}. We caution that this fiducial distribution
  function is only for the purpose of testing our method, and the actual distribution
  of galaxies in surveys such as SKA and LSST can be significantly different. \label{fig:ns}}
\efi

In this section we
test the ABS method against the fiducial case which may resemble what
we expect for a stage IV dark energy project.  Testing against more
general cases will be carried out in the next section.

\subsection{The fiducial case}
\label{sub:fiducialsurvey}
Following our previous papers \citep{YangXJ11,YangXJ15}, the fiducial
survey has $N_{\rm tot}=10^9$ galaxies with spectroscopic redshifts
determined by 21cm emission line,
at $0.8<z<1.2$ and over $10^4$ 
deg$^2$. Each of these flux bins has $N_{\rm
  tot}/N_F=2\times 10^8$ galaxies. The galaxy luminosity function,
along with the range of flux bins,  are shown in Fig. \ref{fig:ns}. The fiducial sky coverage is
$f_{\rm sky}=10^4/(4\pi/(\pi/180)^2)$. As shown in \citet{YangXJ11},
this can be achieved by a SKA like 
radio array of total collecting area of 1 km$^2$, through 21cm observation of
neutral hydrogen in galaxies. This requires the full size phase-2 SKA with
configuration optimized for a dedicated 5 year 21cm survey. Although
this is not likely happening for SKA,  it demonstrates that such
requirement on galaxy surveys is indeed within the capability of
future surveys. Therefore it serves as a suitable example for the purpose
of this paper. Furthermore, imaging surveys such as LSST will have
billions of galaxies with good photometric redshift measurements, and
will be comparable to the fiducial survey in lensing reconstruction
through cosmic magnification.  

%%%%%%%%%%
\bfi{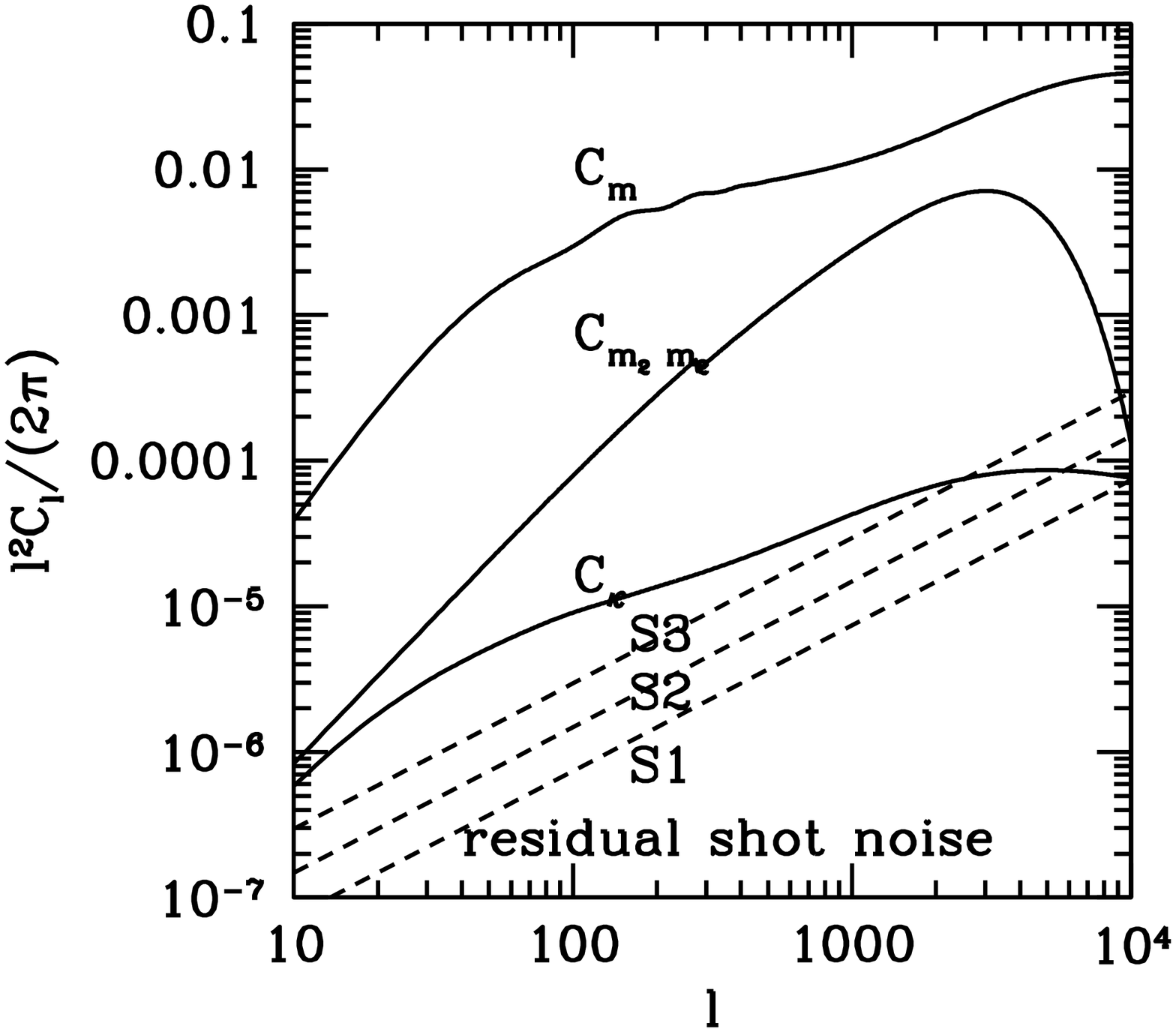}
\caption{Contaminations of lensing measurement by cosmic magnification
  (magnification bias), at source redshift $0.8<z<1.2$. The measured
  galaxy clustering power spectrum is the sum of the lensing signal
  ($\sim C_\kappa$) and the galaxy intrinsic clustering ($\sim
  C_{m},C_{m_2m_2}$), with measurement error caused by shot noise
  (dash lines, with bin size $\Delta \ell=0.4\ell$). The
  bottom dash line with label ``S1'' is that of the fiducial survey, 
  and the other two lines with label ``S2'' and ``S4'' have a factor
  of $2$ and $4$ larger error, respectively.   \label{fig:cl}}
\efi
%%%%%%%%%%%

For each fixed $\ell$, the intrinsic galaxy power spectra $C^g_{ij}$
($i,j=1,\cdots, N_F$) is a real, positive definite,  and symmetric matrix.  According to the spectral
decomposition theorem, it  can always be
decomposed into
\ba
\label{eqn:CgV}
C^g_{ij}=\sum_{\alpha=1}^M C_\alpha V^{(\alpha)}_i
V^{(\alpha)}_j\ .
\ea
Here, ${\bf V}^{(\alpha)}$ is the $\alpha$-th eigenvector of the
matrix ${\bf C}^g$. $C_\alpha$ is the corresponding
eigenvalue, which is real and positive. \citet{Bonoli09,Hamaus10}
found that there are two to three eigenmodes in $C^g_{ij}$. This motives us to adopt the following fiducial model of the intrinsic galaxy clustering,
\be
\label{eqn:cg}
C^g_{ij}=b^{(1)}_ib^{(1)}_jC_m+b^{(2)}_ib^{(2)}_jC_{m_2m_2}\ . \ee
Details of the two biases ($b^{(1,2)}$) and caveats of the model are
given in the appendix \ref{sec:bias}.  Although not exact, throughout the paper we call
$b^{(1)}$ as the linear bias and $b^{(2)}$ as the quadratic bias for
convenience. Furthermore, we will approximate $b^{(1)}$ as the
deterministic bias shown in Eq. \ref{eqn:CL}. $C_m$ is the 
angular power spectrum of the projected matter density
field. $C_{m_2m_2}$ is the power spectrum of 
projected $\delta^2_{m,R}$, where $\delta_{m,R}$ is the matter
density smoothed over a radius $R=1\mpch$. 

%%%%%%%%
\bfi{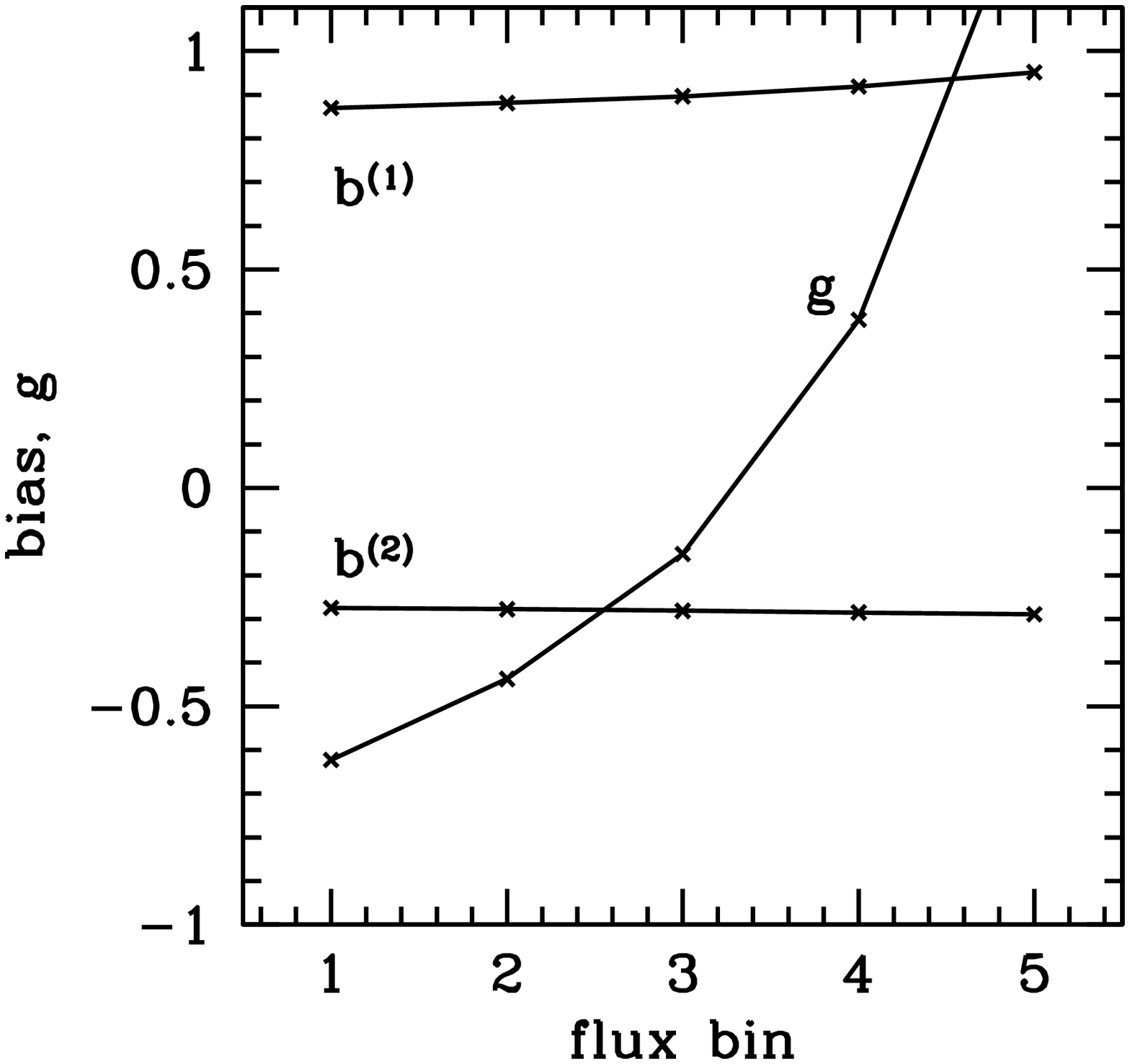}
\caption{The fiducial galaxy biases and $g$ in the $5$ flux
  bins. $b^{(1)}$ is the linear bias and $b^{(2)}$ is the
  quadratic bias. The prefactor $g$ in the magnification bias changes
  sign from faint end to bright end.  \label{fig:bg}}
\efi
%%%%%%%%%%%

Fig. \ref{fig:cl} shows $C_m$, $C_{m_2m_2}$ and
$C_{\kappa}$ for galaxies at $0.8<z<1.2$. As expected,  the lensing
signal $C_\kappa$ is a factor of $10^2$ to $10^3$ smaller than the galaxy
intrinsic clustering. The dominant contribution of galaxy intrinsic
clustering comes from $b^{(1)}$. But the contribution
from $b^{(2)}$ is also significantly larger than the lensing
signal.  Fig. \ref{fig:cl} also shows  $\sigma_{\rm shot}$, the r.m.s of the residual
shot noise of the fiducial survey (the bottom dash line labelled with
``S1''). It is significantly smaller than the lensing signal
$C_\kappa$ that we aim to reconstruct in the range $\ell\la
5000$. Later in \S \ref{subsec:test} we find that this leads to
excellent reconstruction of $C_\kappa$ at $\ell\la 5000$. The dash
line with label ``S2'' has $\sigma_{\rm shot}$ enlarged by a factor
of $2$ and the one with label ``S3'' enlarged by a factor of $4$. The
ABS performance against these $\sigma_{\rm shot}$ will be carried out
in \S \ref{sec:test2}.

The necessary condition to extract the lensing signal (cosmic magnification) from
galaxy intrinsic clustering is that
the flux dependence of the lensing  signal differs from that of the galaxy intrinsic
clustering. Namely, in the $N_F$ dimension flux space, the vector ${\bf g}$ must not
be parallel to the vectors ${\bf b}^{1,2}$. Fig. \ref{fig:bg} shows the results of $b^{(1,2)}_i$,
and $g_i$ ($i=1,\cdots, N_F$). Clearly, the flux dependence of $g$ is
significantly different to that of biases. It not only has a stronger
dependence, but also changes sign, from negative at faint end to
positive at bright end. Unless by coincidence, we expect no
eigenvector ${\bf v}^{(\alpha)}$ in Eq. \ref{eqn:CgV} to $\propto
{\bf g}$. Therefore we believe that  unbiased lensing
reconstruction through cosmic magnification is always doable
(Eq. \ref{eqn:analyticalsolution}), as long as the galaxy survey is
sufficiently powerful such that  statistical error in the galaxy power
spectra measurement is sufficiently small. 

\subsection{Test results}
\label{subsec:test}
\bfi{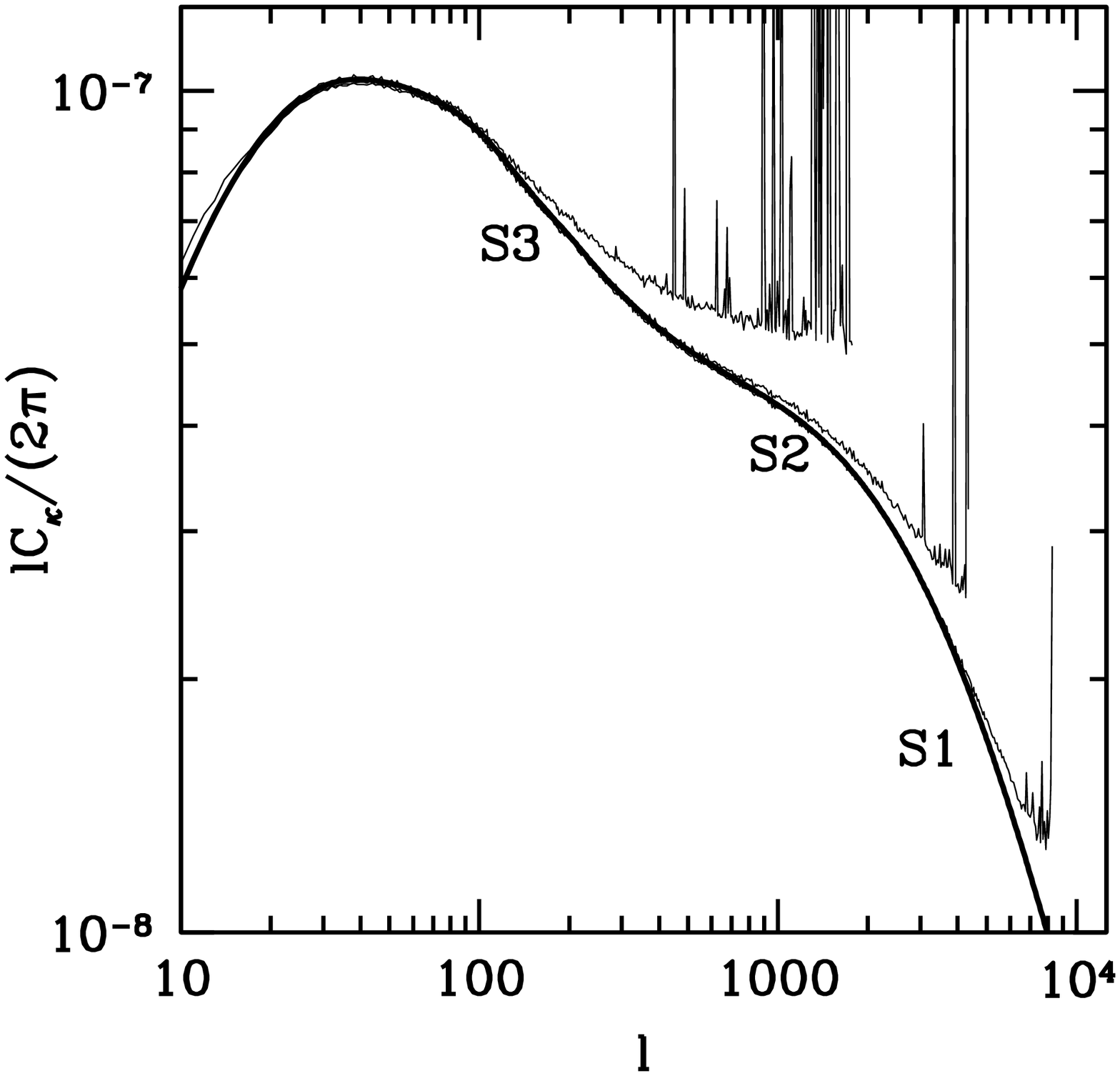}
\caption{The accuracy of  lensing power spectrum reconstruction, for
  the three cases of residual shot noise in Fig. \ref{fig:cl}. The
  thick solid line is the input $\tilde{C}_\kappa$ and the thin lines
  are the output by our ABS method, averaged over $1000$ realizations
  of residual shot noise. The reconstruction fails when the
  residual shot noise is comparable to the lensing signal
  ($\sigma_{\rm shot}\sim C_{\kappa}$), with more exact
  dependence given by the $\lambda_\mu$-$c_\mu$ plot in Fig. \ref{fig:c}.  \label{fig:n}}
\efi

We  assume Gaussian shot noise error $\delta C_{ij}^{\rm shot}$, with
r.m.s. dispersion given by as Eq. \ref{eqn:sigmashot}.  We choose a
relatively large bin size $\Delta 
\ell=0.4\ell$, to suppress shot noise.  For each $\delta C_{ij}^{\rm shot}$, we generate $1000$
realizations for the tests.  Such realizations are then added to $C^L$ of
Eq. \ref{eqn:CL} to generated 1000 realizations of simulated $C^{\rm
  obs}_{ij}$ using Eq. \ref{eqn:CLobs}.  We then use the ABS method to
process the simulated $C^{\rm obs}_{ij}$ and output
$\tilde{C}_\kappa$.

The test result is shown in
Fig. \ref{fig:n}. The thick solid line is the input
$\tilde{C}_{\kappa}$.\footnote{Notice again that for $0.8<z<1.2$,
$r^2_{m\kappa}\simeq 0.01$, so $\tilde{C}_\kappa$ is $1\%$ smaller
than $C_\kappa$. Although the difference is small, it is often
comparable to the systematic error of weak lensing
reconstruction. Therefore we have to keep it in mind when testing our
method.} The thin solid line with label ``S1''  is the output by our
ABS method applied to the fiducial survey. It shows that, our ABS method successfully recovers the
input $\tilde{C}_\kappa$ up to $\ell\sim 5000$. Towards smaller scale,
it begins to overestimate the lensing power spectrum and the
performance quickly degrades. It eventually becomes unstable at
$\ell\ga 7000$. This is caused by the increasing 
fluctuations of residual shot noise with respect to the lensing signal
(Fig. \ref{fig:cl}).  Fig. \ref{fig:n} also shows the test results
against the cases of ``S2'' and ``S3'' with a factor of $2$ and $4$
larger $\sigma_{\rm shot}$. Clearly the reconstruction is highly
sensitive to $\sigma_{\rm shot}$. We postpone detailed discussion on
such dependence until in \S \ref{sec:test2}. 

To highlight the reconstruction accuracy, we plot the fractional 
statistical and systematic error (solid lines) in 
Fig. \ref{fig:cut}. Both errors are evaluated averaging over $1000$
realizations of residual shot noise.  At $\ell\la 1000$, the systematic error is negative, with a
relative amplitude of  $\sim 1\%$. However at $\ell\ga 2000$, it
becomes positive and reaches $\sim 10\%$ at $\ell=5000$. To understand
such behavior of the systematic error, we will resort to the 
$\lambda_\mu$-$c_\mu$ diagnostic in \S \ref{subsec:lambdamucmu}. 
Nevertheless, at $\ell\la 7000$, the
systematic error is smaller than the statistical error. Namely, despite the
existence of systematic errors, the lensing reconstruction is {\it statistically}
unbiased.

%%%%%%%%%%%%
\bfi{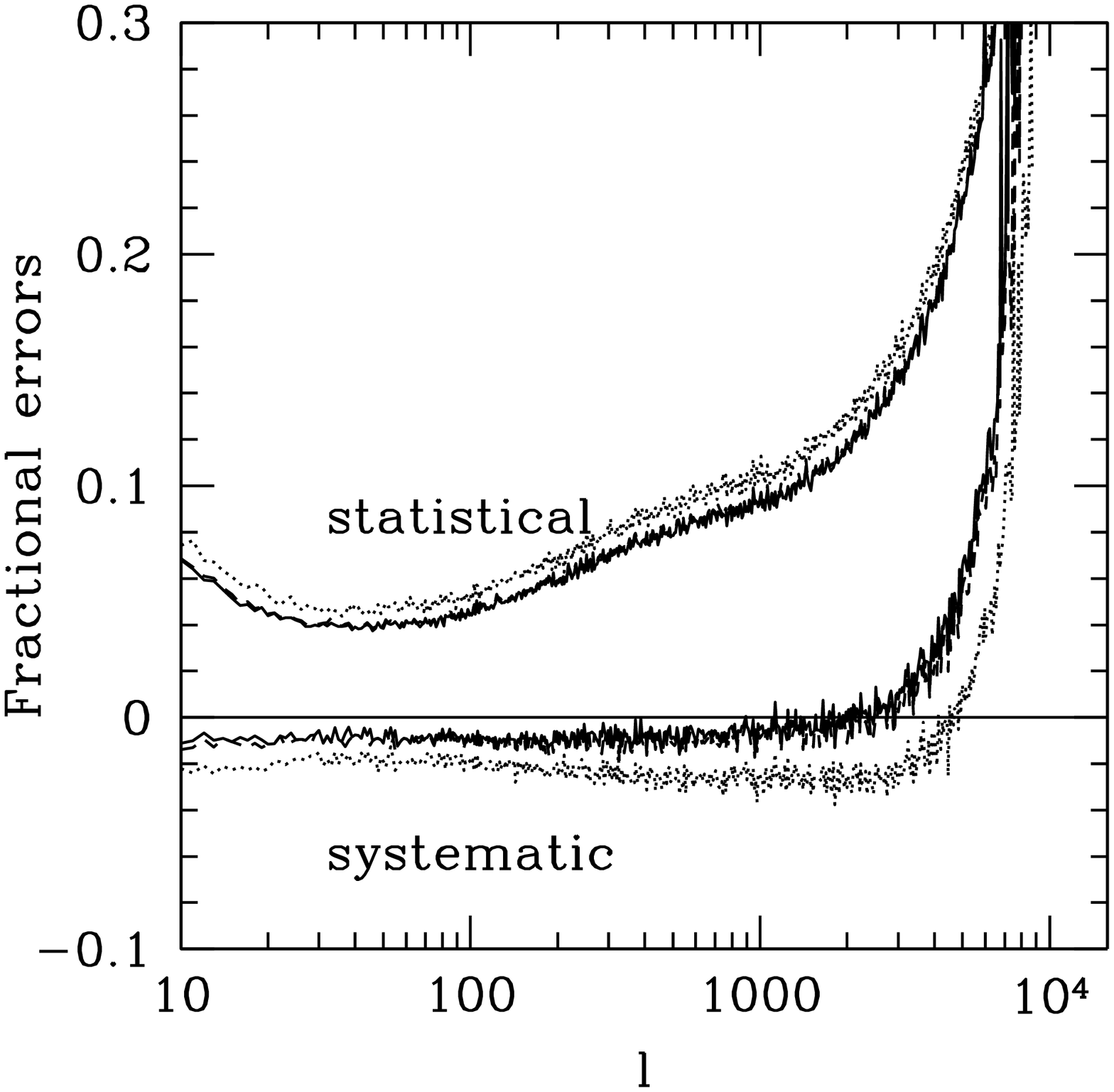}
\caption{The dependence of lensing power spectrum determination on the
  cut of eigenvalue $\lambda_{\rm cut}$. To highlight the dependences,
  we show both the relative statistical
  error and relative systematic error for $\lambda_{\rm cut}=\sigma_{\rm
    shot}$ (solid lines), $\sigma_{\rm shot}/N_F^{1/2}$ (dot lines)
  and $2\sigma_{\rm shot}/N_F^{1/2}$ (dash lines, almost overlap with
  solid lines). Both statistical and systematic error are estimated
  from $1000$ realizations of simulated data. \label{fig:cut}}
\efi
%%%%%%%%%%%%

\subsection{The $\lambda_\mu$-$c_\mu$ diagnostic for the fiducial
  case}
\label{subsec:lambdamucmu}
The fiducial case has three physical eigenmodes, since there
are three independent vectors in the $N_F$ dimension flux space (${\bf
  b}^{(1,2)}$ and ${\bf g}$). The eigenmodes depend on the multipole
$\ell$. Since the lensing power spectrum suffers from non-negligible uncertainties of baryon
physics at $\ell\ga 2000$
\citep{2004APh....22..211W,2004ApJ...616L..75Z,2006ApJ...640L.119J,2008ApJ...672...19R},
$\ell\la 2000$ is most useful for
weak lensing cosmology. Therefore we show the result of $\ell=2000$
(filled circles) in
Fig. \ref{fig:c}.  Since $g$ is negative at faint end and
positive at bright end, $\sum g_i\sim 0$. Since both $b^{(1)}$ and
$b^{(2)}$ vary slowly with flux, the two vectors ${\bf b}^{(1,2)}$ are
nearly orthogonal to ${\bf g}$ (${\bf b}^{(1,2)}\cdot {\bf g}\sim
0$). Therefore one eigenvector (the second one) is almost perfectly parallel to ${\bf
  g}$, with $c$ very close to unity. The majority of the intrinsic
clustering is absorbed in the first (the largest) eigenmode. Its eigenvalue is $\sim
\sum_i C^g_{ii} $, more than two orders of magnitude larger than the
second eigenvalue. The ABS method automatically suppresses this
eigenmode in the lensing reconstruction, through a small $G_1 \ll 1$
and through $\lambda_1\gg 
C_\kappa$. For these reasons,  its contribution ($c_1$) is less than $0.01\%$. The rest of
the intrinsic clustering and the lensing signal enter the third
eigenmode. Its eigenvalue is too tiny to be detected in realistic
surveys. But since it is almost completely orthogonal to ${\bf g}$
($G_3\simeq 0$), its contribution is also negligible. 

%%%%%%%%
\bfi{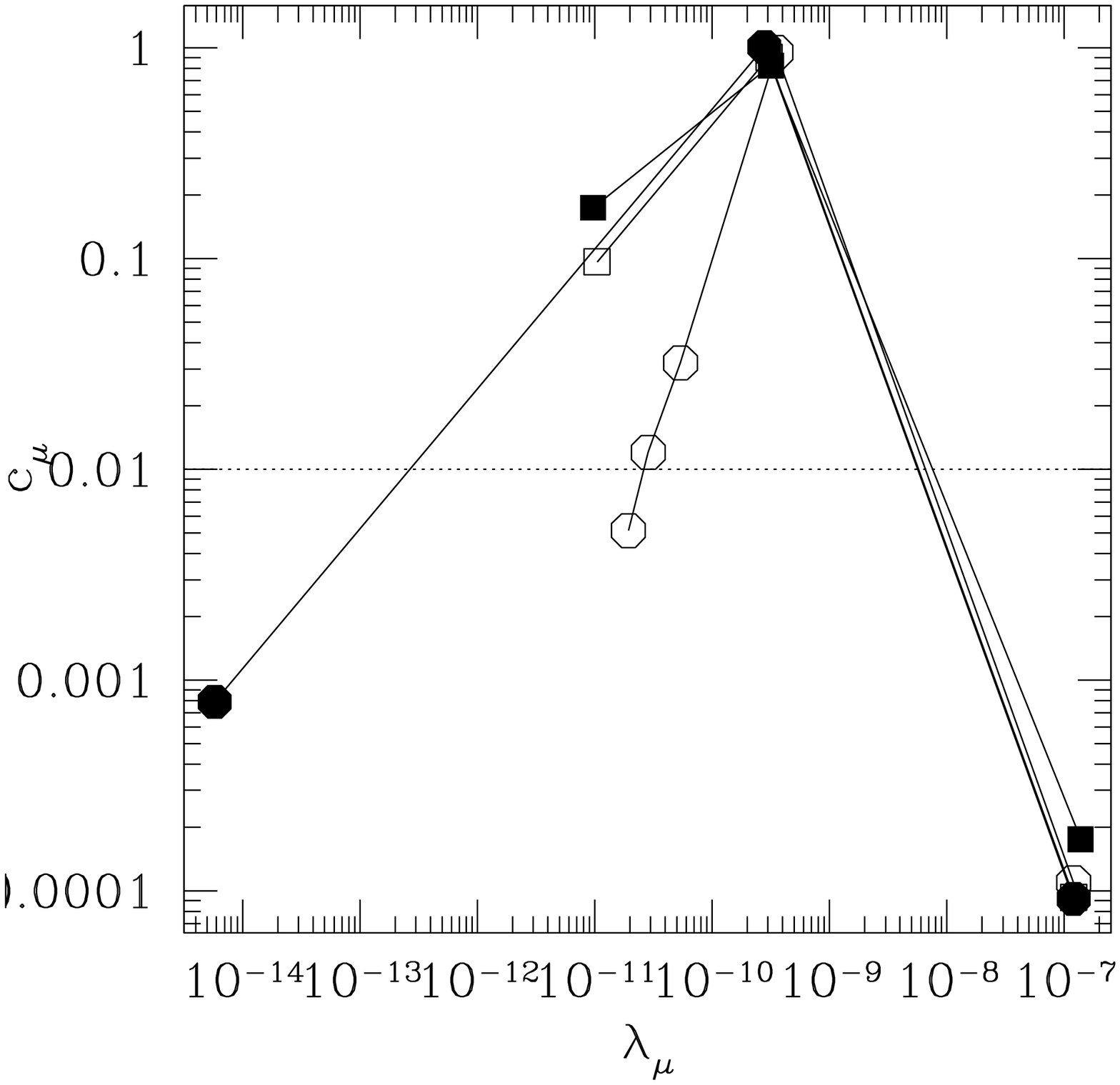}
\caption{The $\lambda_\mu$-$c_\mu$ plot at $\ell=2000$. $c_\mu$ is the contribution
  of the $\mu$-th physical eigenmode to the lensing power spectrum reconstruction. We show four cases, the
  fiducial case (filled circle),  shape of ${\bf b}^{(1,2)}$ changing by
  $30\%$ (filled square and open square respectively) described in \S \ref{subsec:biass}, 
  and a more complicated case (open circle) described in \S
  \ref{subsec:systematic}.  For
  the reconstruction to be accurate, all physical eigenmodes
  with significant $c_\mu$ must be robustly identified
  ($\lambda_\mu\gg \sigma_{\rm shot}$).\label{fig:c}}
\efi
%%%%%%%%%%

Therefore, there is only one eigenmode (the second eigenmode) relevant to the lensing
reconstruction. The corresponding eigenvalue is $\sim
\sum_i^{N_F} g_i^2 C_\kappa\sim C_\kappa$.  In this case,  the condition for accurate lensing reconstruction
is $\sigma_{\rm shot}\ll C_\kappa$. Fig. \ref{fig:cl} shows that this
condition is well satisfied at $\ell\leq 1000$ and roughly satisfied
at $\ell \la 5000$. Therefore our ABS method accurately recovers the
input lensing signal.  At $\ell\ga 5000$, this condition is no longer
satisfied, the second physical eigenmode is not robustly detected, and
the reconstruction suffers from non-negligible systematic bias. At
$\ell\ga 10^4$, the second physical eigenmode is overwhelmed by
residual shot noise. The reconstruction is highly unstable and no longer reliable.   

Fig. \ref{fig:c} also shows $\lambda_\mu$-$c_\mu$ for three other
cases, corresponding to different biases and different clustering
components. The distribution of $\lambda_\mu$-$c_\mu$ varies
significantly among these cases. Later in \S \ref{sec:test} we will
discuss their impacts on the lensing reconstruction. 

\subsection{The choice of $\lambda_{\rm cut}$}
\label{subsec:cut}
The above $\lambda_\mu$-$c_\mu$ diagnostic also tells us that the
choice of $\lambda_{\rm cut}$ is important. $\lambda_{\rm cut}$ has
two-fold and competing impacts on the reconstruction. On one
hand, it serves to exclude unphysical eigenmodes caused by statistical
measurement errors (residual shot noise in this paper). If it fails
and one unphysical eigenmode is wrongly included,  the reconstructed
$\tilde{C}_\kappa$ will be underestimated. For this purpose,  the
larger $\lambda_{\rm cut}$, the better. On other hand, 
the cut may also wrongly exclude physical eigenmodes and cause
overestimation of $\tilde{C}_\kappa$. To alleviate this problem, the
smaller $\lambda_{\rm cut}$ the better. We try a few choices of $\lambda_{\rm cut}$ to
demonstrate its impact. 

Since we know that $\lambda_{\rm cut}$ should be comparable to $\sigma_{\rm
  shot}$, we try three possibilities, $\lambda_{\rm cut}=\sigma_{\rm
  shot},\sigma_{\rm shot}/\sqrt{N_F},2\sigma_{\rm
  shot}/\sqrt{N_F}$. The results for $\lambda_{\rm cut}=\sigma_{\rm
  shot}$ and $2\sigma_{\rm
  shot}/\sqrt{N_F}$ are almost identical(Fig. \ref{fig:cut}). The smallest cut
($\lambda_{\rm cut}=\sigma_{\rm shot}/\sqrt{N_F}$) is more efficient
to include physical eigenmodes in the summation of
Eq. \ref{eqn:analyticalsolution} and thus reduces/improves the overestimation
at $\ell\ga 5000$. The price to pay is the higher probability of wrongly
including unphysical eigenmodes, and thus stronger/worse
underestimation at $\ell\la 2000$.  If we focus on the measurement at $\ell<2000$ of greater cosmological importance, the
choice of $\lambda_{\rm cut}=2\sigma_{\rm shot}/\sqrt{N_F}$ or
$\sigma_{\rm shot}$ are both reasonable. Hereafter we will adopt
$\lambda_{\rm cut}=2\sigma_{\rm shot}/\sqrt{N_F}$, since it takes the
dependence of noise eigenvalue on the number of flux bins into account. 

A remaining task is to fix the optimal $\lambda_{\rm
  cut}$ minimizing the systematic error of $\tilde{C}_\kappa$
determination. A related question is that whether we shall go beyond
the step function. For example, given the noise properties (Eq. \ref{eqn:shotnoise}),
one can estimate the probability of a given eigenvalue to be
contaminated by measurement noise. This may allow the design of an
optimal weighting function of $\lambda$, instead of a step
function used in this paper. This is an important
issue for further investigation.

\section{Testing against more general cases}
\label{sec:test2}
The excellent performance of our ABS method to reconstruct the lensing
power spectrum for the fiducial case is exciting. It demonstrates its
great potential for stage IV dark energy surveys such as LSST and
SKA. However, the  performance of this method may depend on many
factors, for which the fiducial case may no longer be
representative. These factors include (1) survey specifications which
determine $\sigma_{\rm shot}$, (2) galaxy biases and (3) systematic
errors in the measured galaxy power spectra. In future  we aim to fix these uncertainties by
targeting at a specific survey. By targeting at a specific survey and
calibrating against N-body simulations, we may fix these factors and
make more specific  forecast. We will carry out this exercise in a
companion paper. 

Instead of being specific, here  in this methodological paper we will take an alternative
approach. We explore a wide range of possibilities, by varying
$\sigma_{\rm shot}$ that a survey can achieve, the amplitude/shape of biases, and extra bias
components/systematic errors in the galaxy power spectrum
measurement. Since there are so many
dimensions of possibilities, we  can not fully probe the parameter
space. Instead, each time we just  vary one configuration and fix the
rest as the fiducial ones.These tests   demonstrate the generality of the ABS method.  

%%%%%%%%%%%%%%%%
\bfi{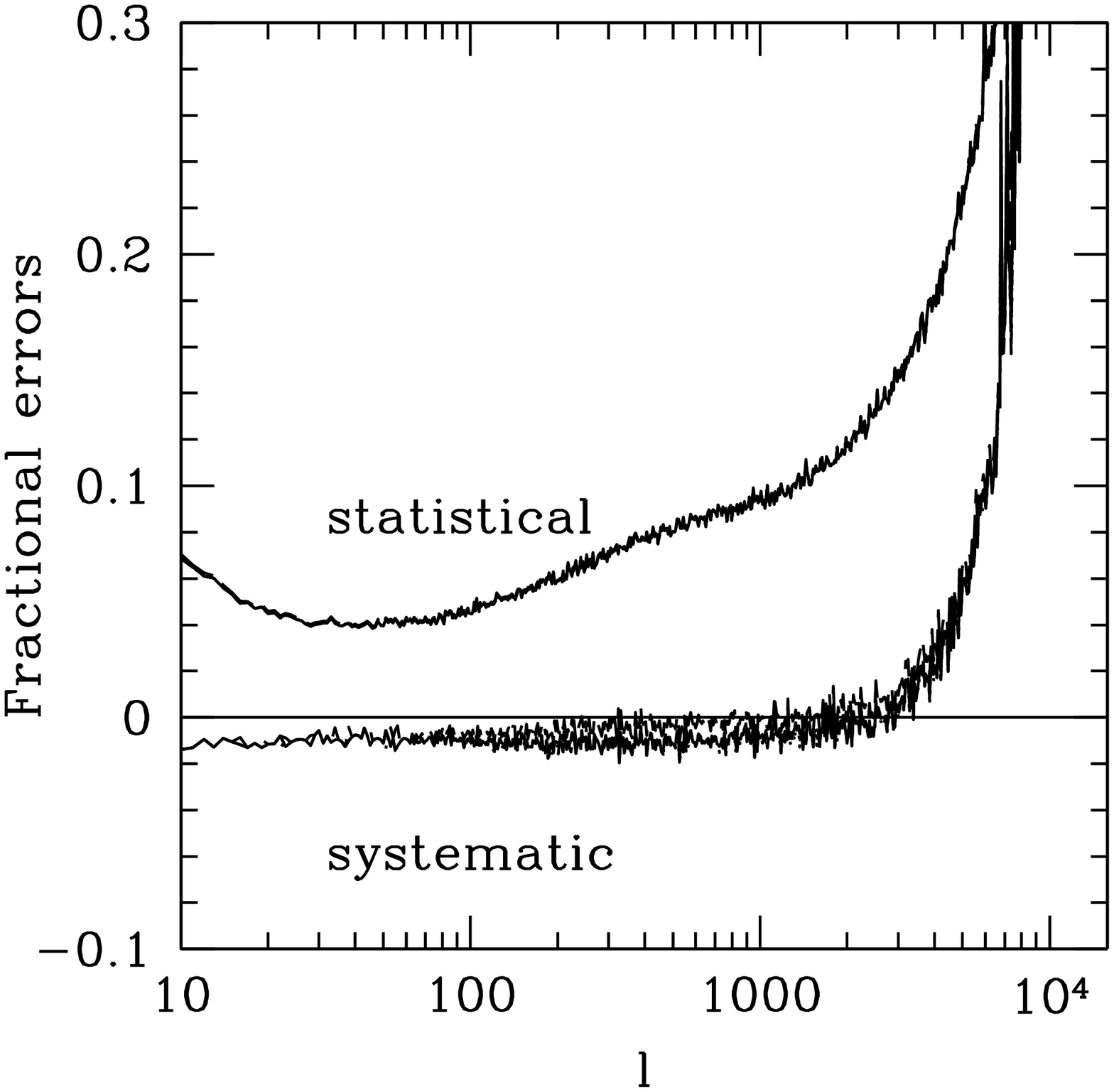}
\caption{The dependence of lensing power spectrum determination on the
  amplitude of the linear bias ${\bf b}^{(1)}$. We arbitrarily increase the
  amplitude of ${\bf b}^{(1)}$ by a factor of $2$ and $4$. Neither the
  statistical nor the systematic error change visibly. However, if we
  reduce ${\bf b}^{(1)}$ by a factor of $10$ (dash lines), the systematic error
  has a small but visible change.  \label{fig:b1}}
\efi
%%%%%%%%%%%%%%%%

\subsection{The impact of survey specifications}
\label{subsec:survey}
The most important factor determining the accuracy of weak lensing
reconstruction  is the measurement error in the galaxy power
spectra. In particular, shot noise fluctuations can bias the lensing
reconstruction, as discussed in detail by the $\lambda_\mu$-$c_\mu$
diagnostic. 

Fig. \ref{fig:n} shows that the lensing reconstruction strongly depends on $\sigma_{\rm shot}$.  For the fiducial one, we are able to reconstruct
the lensing power spectrum accurately up to $\ell \sim 5000$. But a
factor of $2$ increase in $\sigma_{\rm shot}$ renders the accurate
measurement at $\ell \ga 2000$ impossible. A further factor of $2$
increase renders the accurate measurement at $\ell \ga 300$
impossible.  This strong dependence on $\sigma_{\rm shot}$ comes from
the fact that we need the condition $g^2C_\kappa\gg \sigma_{\rm shot}$ so that the
lensing eigenmode can be robustly identified. From
Fig. \ref{fig:bg}, $g^2\equiv \sum_i g_i^2=2.7$. Let us focus on
$\ell=1000$, where
$C_\kappa\simeq 5\sigma_{\rm shot,fid}$. Therefore, the above
condition  is well satisfied by the fiducial case (``S1'') and the
lensing reconstruction at $\ell=1000$ is accurate. But it is not
satisfied by ``S3'',  resulting in failure of lensing reconstruction.
The case of ``S2'' falls somewhere between the two.

$\sigma_{\rm shot}$ is set by survey specifications such as the total
number of galaxies $N_{\rm tot}$ and the sky coverage $f_{\rm sky}$. Since $\sigma_{\rm
  shot}\propto f_{\rm sky}^{1/2}/N_{\rm tot}\propto
\bar{n}^{-1}f^{-1/2}_{\rm sky}$,  successful reconstruction requires
sufficiently high galaxy number density $\bar{n}$ and sufficiently large $f_{\rm sky}$. The fiducial value
of $\sigma_{\rm shot}$ adopted is achieved for a survey with
high galaxy number density  ($\bar{n}=28$ galaxies per arcmin$^2$ at $0.8<z<1.2$),
and large sky coverage ($f_{\rm sky}=0.24$). The survey requirements
are stringent, but within the capability of SKA
\citep{YangXJ11,YangXJ15}. 

These requirements may also be satisfied by
other surveys. (1) SKA has a proposal to survey
$30000$ deg$^2$ for the detection of $9\times 10^8$ HI galaxies.\footnote{https://pos.sissa.it/archive/conferences/215/017/AASKA14\_017.pdf} However, these are galaxies above
$10\sigma$ detection threshold. For cosmic magnification measurement,
we can use fainter galaxies (e.g. at $5\sigma$), and there are many
more of them.  Furthermore, for fixing total survey time, smaller sky
coverage results in higher  galaxy number density. So there is room to
improve the cosmic magnification  measurement by adjusting the SKA survey strategy. 
(2) Imaging surveys with reasonably good photo-z can also be used for
lensing reconstruction through cosmic magnification. Since we
do not  need to measure galaxy shapes, we do not require galaxies to
be as bright and large as in cosmic shear measurement. Therefore for
the same imaging survey, there will be significantly more galaxies 
useful for lensing magnification measurement than for shear
measurement. One possibility is LSST. LSST will have $\sim 30$ galaxies per arcimin$^2$ for
cosmic shear measurement. Therefore it will have many more galaxies
for the cosmic magnification measurement. One open question is
  the extra error induced by photo-z errors, which affect both the
  signal through the determination of $g$, and the intrinsic
  clustering. This is to be studied in detail
against N-body mocks in the future. (3) Furthermore, we may even
choose wider redshift bin size or even use galaxies over  the whole survey
redshift range to perform the lensing reconstruction. In this case,
the SKA galaxy radio continuum survey is
also a good target. It is expected to detect $\sim 5$ billion radio
galaxies over $30000$ square
degrees.\footnote{https://pos.sissa.it/archive/conferences/215/018/AASKA14\_018.pdf}
Therefore we expect that we may at least realize the case of ``S2'',
possibly ``S1'', and likely even better. 

%%%%%%%%%%%
\bfi{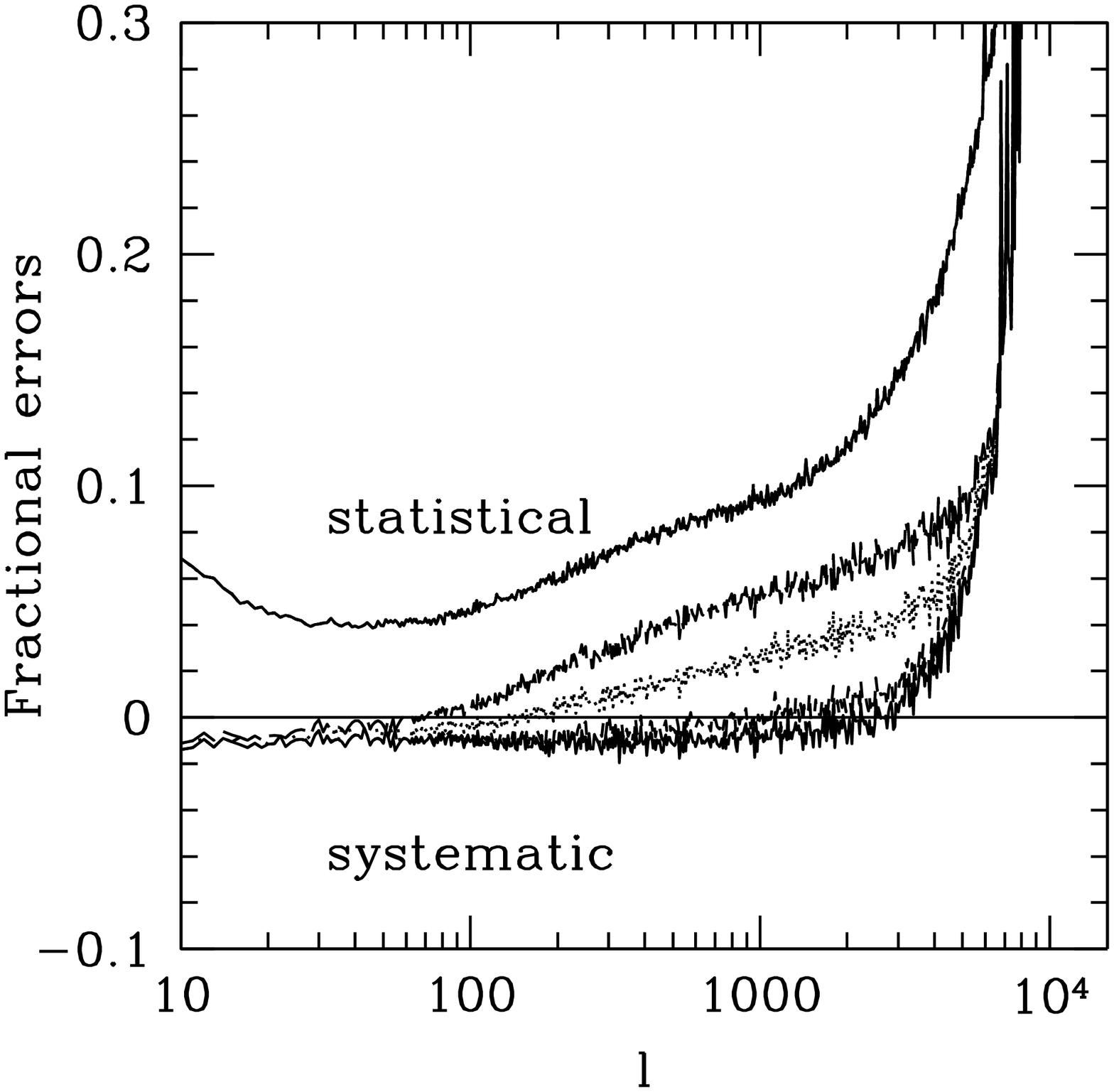}
\caption{The dependence of lensing power spectrum determination on the
  amplitude of the quadratic bias (${\bf b}^{(2)}$.  We enlarge ${\bf
    b}^{(2)}$ by a factor of $2$ (dash lines), $4$ (dot lines) and $6$
  (long dash lines). The statistical error has negligible dependence on its amplitude. But the
  systematic error is sensitive to its amplitude and the resulting
  stochasticity.  \label{fig:b2}}
\efi
%%%%%%%%%%%

\subsection{The impacts of galaxy biases}
\label{subsec:bias}
The galaxy intrinsic clustering is the major systematic contamination of weak lensing
reconstruction through cosmic magnification. At the level of two-point
statistics, it is completely fixed by the galaxy biases (linear,
quadratic, etc.). The amplitude and the shape (flux dependence) of
galaxy biases affect the lensing reconstruction differently. Therefore we investigate the two separately.

\subsubsection{The impacts of  galaxy bias amplitude}
To isolate the impact of galaxy bias amplitude, we fix the shape of
galaxy biases. We first arbitrarily enlarge the amplitude of ${\bf
  b}^{(1)}$ by a factor of $2$, $4$ and $10$. Fig. \ref{fig:b1} shows
the corresponding statistical and systematic errors. Surprisingly,
both errors show negligible dependences on the amplitude of ${\bf
  b}^{(1)}$.  Interestingly, when  we arbitrarily decreases ${\bf b}^{(1)}$ by a
factor of $10$, we find small but visible change in
the systematic error. Even so, the statistical error remains
unchanged.  The insensitivity on the amplitude of ${\bf b}^{(1)}$ has a strong
physical origin. We rely on the different flux dependences to separate
the magnification bias from intrinsic clustering. Roughly speaking, we
compare the difference in $\delta_g$ between bright and faint
samples to extract/reconstruct the lensing signal. Therefore the
lensing reconstruction is insensitive to the overall amplitude of
${\bf b}^{(1)}$, as long as it is the dominant part of galaxy biases. 

We also enlarge ${\bf b}^{(2)}$ by a factor of $2$, $4$ and $6$
to test our method (Fig. \ref{fig:b2}). Now the systematic error is
sensitive to the amplitude of ${\bf b}^{(2)}$. The amplitude of ${\bf
  b}^{(2)}$ determines the stochasticity of intrinsic galaxy
clustering. The stochasticity causes decorrelation between the galaxy
overdensities of different flux bins. It then reduces the efficiency
of extracting lensing signal by comparing between bright and faint
galaxies.  In contrast, we find surprisingly that the statistical error remains
insensitive to the amplitude of ${\bf b}^{(2)}$.\footnote{So far we do not
have a solid explanation on this behavior. We suspect, but without a
proof,  that the
prefactor $\eta$ in Eq. \ref{eqn:error} 
is invariant under the transformation ${\bf b}^{(1)}\rightarrow A{\bf
  b}^{(1)}$ (or ${\bf b}^{(2)}\rightarrow A{\bf b}^{(2)}$)
where $A$ is an arbitrary constant.  } 

Therefore we draw the conclusion that the lensing reconstruction
accuracy is sensitive to the  amplitude of $b^{(2)}$, but insensitive
to that of $b^{(1)}$. 

%%%%%%%%%%%%
\bfi{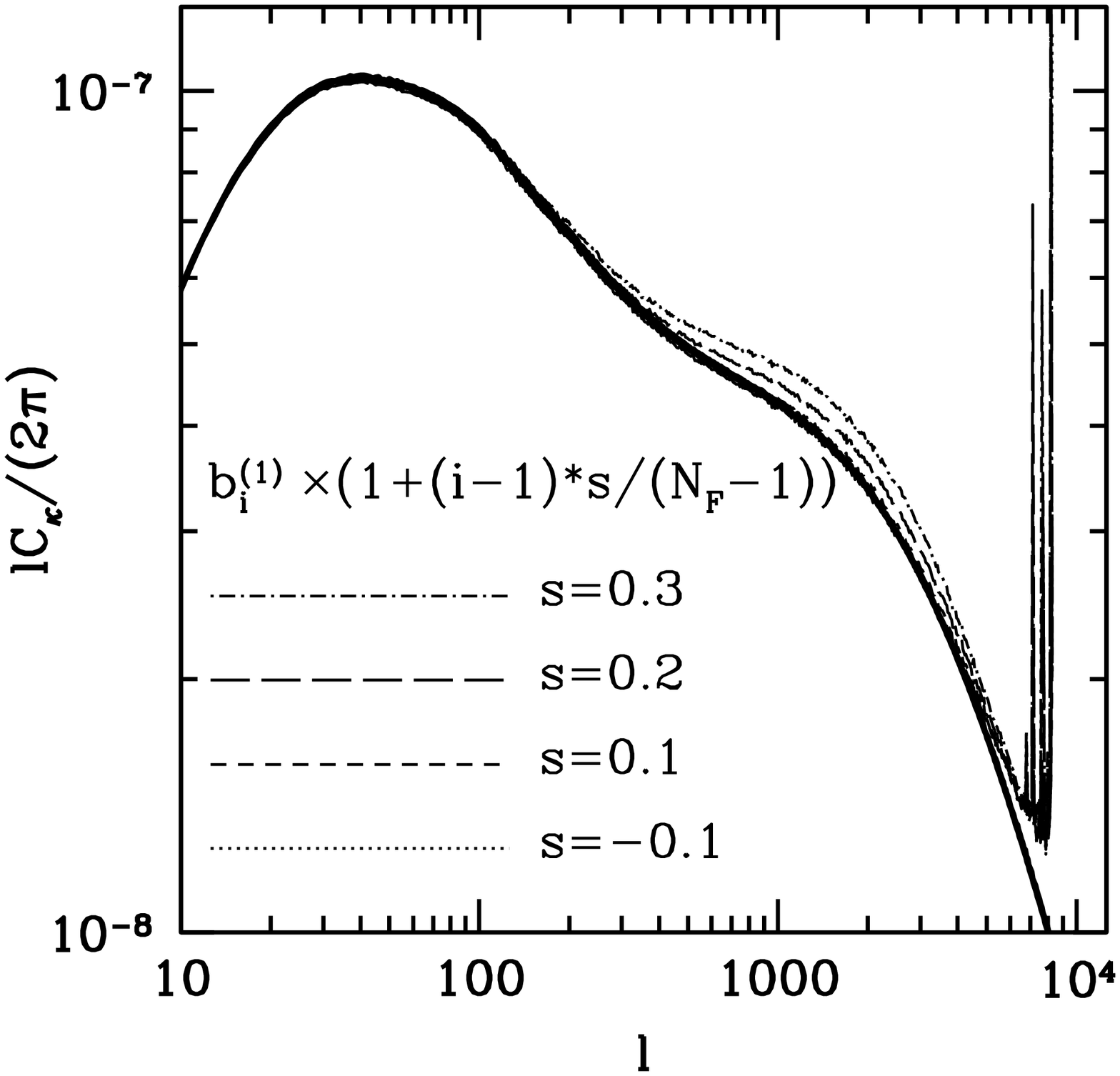}
\caption{The dependence of lensing power spectrum determination on the
  shape of the deterministic bias ${\bf b}^{(1)}$. We change its
  shape from the faint end to the bright end by a factor of $s$.  \label{fig:b1s}}
\efi
%%%%%%%%%%%%
%%%%%%%%%%%%%%
\bfi{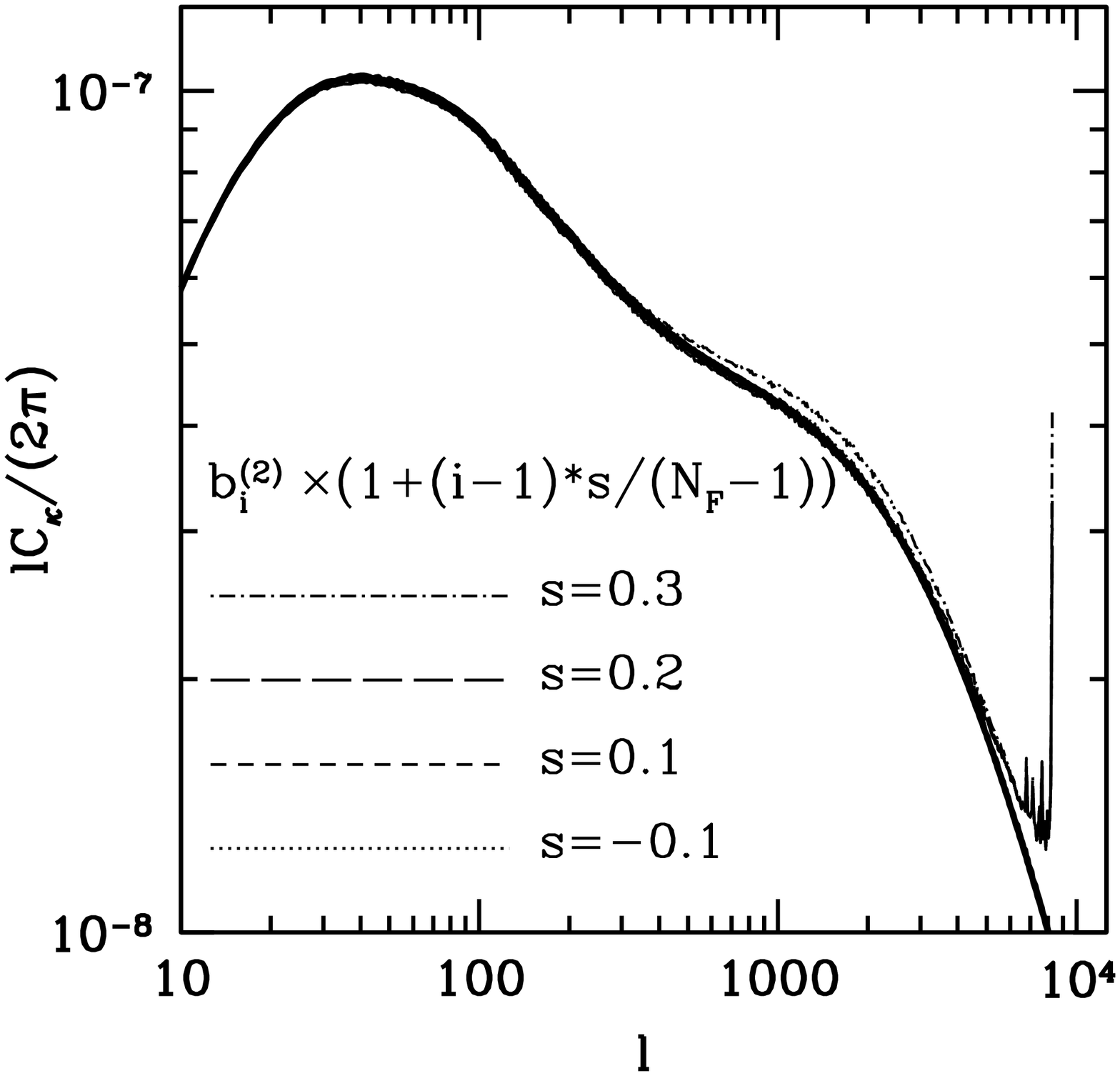}
\caption{The dependence of lensing power spectrum determination on the
  shape of the quadratic bias ${\bf b}^{(2)}$. We change its
  shape from the faint end to the bright end by a factor of $s$.\label{fig:b2s}}
\efi
%%%%%%%%%%%%%%%%
\subsubsection{Impacts of galaxy bias shape}
\label{subsec:biass}
Since we rely on the difference in the flux dependences of the lensing
signal and intrinsic clustering to reconstruct weak lensing, the
lensing reconstruction accuracy is expected to depend on the shape of
galaxy biases in flux space.  We discuss two extreme cases to
demonstrate this dependence. When ${\bf
  b}\parallel {\bf g}$, one can not  separate the galaxy intrinsic
clustering and the lensing magnification 
bias through their flux dependences.   In contrast, when ${\bf
  b}\perp {\bf g}$, the lensing reconstruction would
be most accurate.  For the fiducial intrinsic clustering, ${\bf b}^{(1,2)}$ 
are nearly orthogonal to  ${\bf g}$.  However, this is not completely
coincident. Deep surveys probe not only bright galaxies with positive
$g_i$, but also faint galaxies with negative $g_i$. Since $b_i^{(1)}$
is positive, ${\bf b}^{(1)}\cdot {\bf g}/(gb)$ is expected to be
small.  

To demonstrate the generality of our method, we arbitrarily modify ${\bf b}^{(1)}$ or 
${\bf b}^{(2)}$ by the following recipe
\ba
b_{i}^{(1,2)}\rightarrow b_{i}^{(1,2)}\times \left[1+s\times
  \frac{i-1}{N_F-1}\right]\ .
\ea
A positive $s$ makes the shape of ${\bf b}^{(1)}$ less different to
${\bf g}$ and therefore serves as more stringent test of our ABS
method. We try different values of $s=-0.1,0.1,0.2,0.3$, and the
results are shown in Fig. \ref{fig:b1s}.  As
expected, the reconstruction accuracy depends on the shape of
biases. But even for change as large as $30\%$ in shape, our ABS
method still works to extract the lensing signal. Nevertheless, the systematic error
increases to $\sim 10\%$ at $\ell=2000$.

Fig. \ref{fig:c} shows the eigenmodes of the case $s=0.3$ and the
corresponding contribution factor $c_\mu$, at multipole $\ell=2000$. Since now the shape of
${\bf b}^{(1)}$ is significantly different to ${\bf g}$, it induces 
an extra (the third largest) eigenmode with significant
$c_3=0.17$. When shot noise fluctuation is negligible ($\sigma_{\rm
  shot}\ll \lambda_3$), the reconstruction of $\tilde{C}_\kappa$ will
be unbiased. But when shot noise fluctuation overwhelms  ($\sigma_{\rm
  shot}\gg \lambda_3$), this eigenmode will be completely missing.  It
will then lead to $c_\mu/(1-c_\mu)=20\%$ overestimation of
$\tilde{C}_\kappa$. The actual situation falls between the two
extremes.   The eigenmode has $\lambda_3=9.7\times 
10^{-12}\sim \sigma_{\rm shot}/2$ (Fig. \ref{fig:cl} at $\ell=2000$, but noticing
the $\ell^2/(2\pi)$ prefactor there). The resulting overestimation is
$13\%$ (Fig. \ref{fig:b1s}). 

Fig. \ref{fig:b2s} shows the impact of  ${\bf b}^{(2)}$ shape on
lensing reconstruction. The impact is significantly smaller than that
of ${\bf b}^{(1)}$.  This is again explained by the
$\lambda_\mu$-$c_\mu$ diagnostic. Fig. \ref{fig:c} shows the case of
$s=0.3$ (open square). Changing the shape of ${\bf b}^{(2)}$ by $30\%$
also induces an eigenmode significant for lensing reconstruction, with
$c_3=0.097$. This $c_3$ is a factor of $2$ smaller than the case of
${\bf b}^{(1)}$. This explains the significantly smaller systematic
error, comparing to that of ${\bf b}^{(1)}$. 

Therefore we confirm our expectation that the shape of galaxy bias
(flux dependence) is an important factor in weak lensing
reconstruction. Realistic forecast of lensing reconstruction accuracy
then relies on the reliability of the input fiducial galaxy bias
model.   But we want to emphasize two points. First, the lensing
reconstruction itself does not rely on assumptions of the galaxy bias
model, since we do not fit the data against any bias model. Second,
the systematic error of lensing reconstruction is caused by survey
limitation, instead of fundamental flaw in our method. This systematic error
decreases and eventually vanishes if the galaxy number density is
sufficiently high and the survey area is sufficiently large. 

%%%%%%%%%%
\bfi{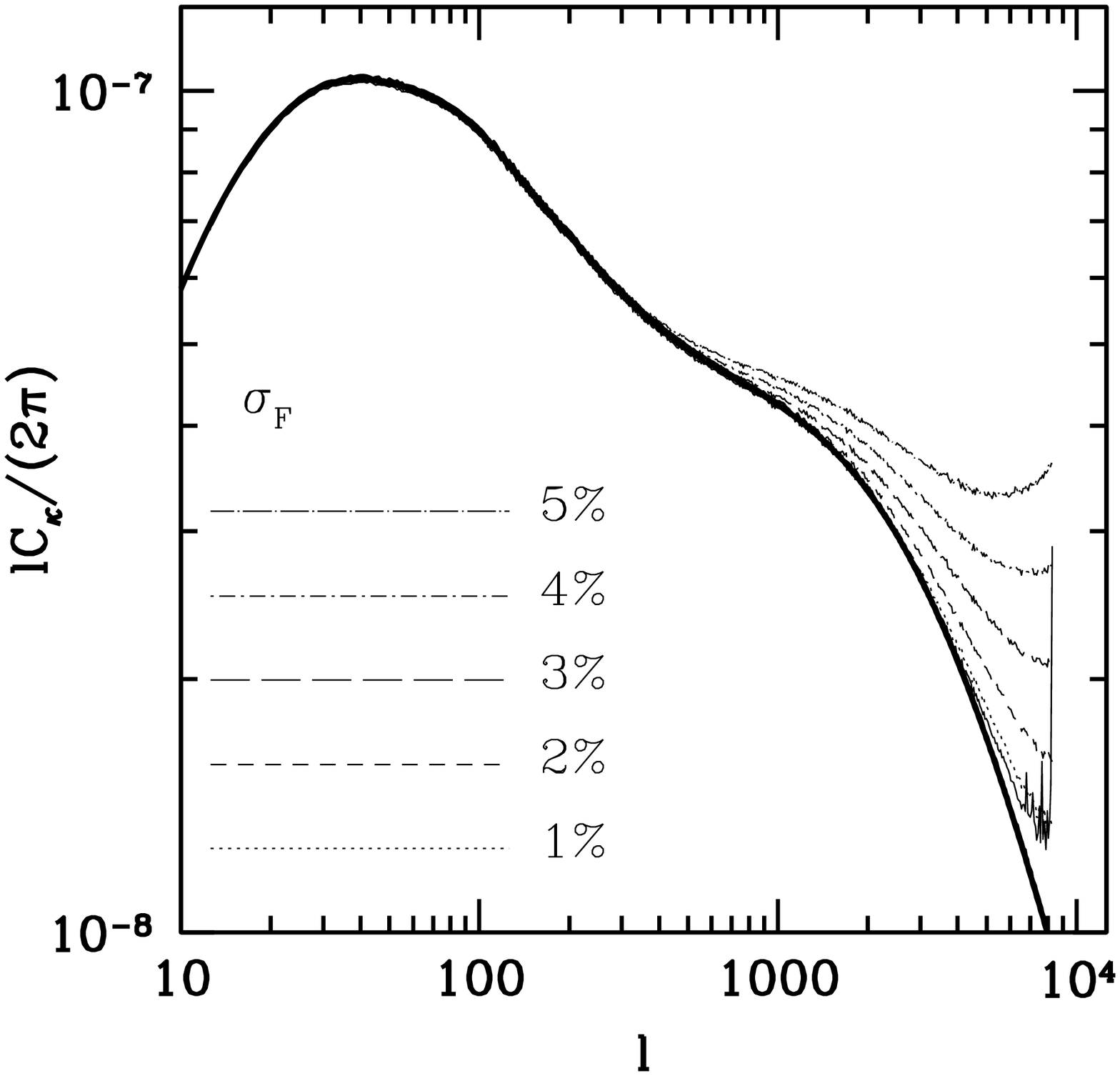}
\caption{The dependence of lensing power spectrum determination on
  extra eigenmodes. By adding $\alpha_i^2 \sigma_F^2/\bar{n}_g$ to
  the diagonal elements of $C^{\rm obs}_{ij}$, we increase the rank
  of the matrix $C^{\rm obs}_{ij}$ to $N_F$, and therefore increase
  the difficulty of lensing reconstruction. Our ABS method still works
  for such extreme test. The observed systematic error is caused by
  observational limitation and will vanish in the limit of
  $\sigma_{\rm shot}\rightarrow 0$.  \label{fig:flux}}
\efi
%%%%%%%%%%
\subsection{Impacts of extra components in the measured $C^{\rm obs}_{ij}$}
\label{subsec:systematic}
The measured matrix $C_{ij}^{\rm obs}$ can have extra principal components than
discussed above. These extra components can  be galaxy biases beyond the linear
and quadratic bias (e.g. \citet{Hamaus10}). They can also be uncorrected/miscorrected errors
in the galaxy clustering measurement. Despite the difference in their physical origins, mathematically speaking they  are similar. They will not only modify existing
eigenmodes, but also generate extra eigenmodes. These extra eigenmodes
increase the difficulty of lensing reconstruction therefore serve as
more stringent tests of our ABS method.

It is beyond the scope of this paper to fully explore these
complexities. Instead, we will focus on a hypothetical case, but with
some physical motivation.
Observationally there are measurement errors in flux. As lensing
magnification, such error in flux measurement also causes
fluctuations in the galaxy number density, $\delta_g\rightarrow
\delta_g+\alpha \delta_F$. Here, $\delta_F\equiv \delta F/F$ is the
fractional flux measurement error.   For random error ($\langle
\delta_F\rangle=0$), on the average it only affects the diagonal
elements $C^{\rm obs}_{ii}$.  Its impact is inversely proportional to $\bar{n}$,
similar to shot noise. In principle, we can predict its ensemble
average and subtract it from the diagonal elements. However, {\it assuming
that it is unnoticed and uncorrected in the measured $C^{\rm
  obs}_{ii}$}, then 
\be
\label{eqn:flux}
C^{\rm obs}_{ij}\rightarrow C^{\rm obs}_{ij}+\left(\alpha^2_i
\frac{\sigma_F^2}{\bar{n}_g}\right)\delta_{ij}\ .
\ee
Here $\sigma_F^2\equiv \langle \delta_F^2\rangle$. With its presence,
the rank of the $N_F\times N_F$ matrix $C^{\rm obs}_{ij}$ will be $N_F$. Instead of
extracting the lensing signal from contaminations of one or two
components, now we need to deal with contaminations of $N_F$ independent
components. Strictly speaking,  the mathematical proof of uniqueness of
$\tilde{C}_\kappa$ reconstruction and the associated analytical
formula presented in this paper no longer hold.  However, even in this
extreme case, our ABS method still works well to some extend. 

Fig. \ref{fig:flux} shows the reconstruction accuracy when such error
exists, for the case of $\sigma_F=1\%$-$5\%$. For $\sigma_F=5\%$, $\tilde{C}_\kappa$ can
still be determined reasonably well. The systematic error is $1.4\%$ at
$\ell=500$, $7\%$ $\ell=1000$, and $19\%$ at
$\ell=2000$.\footnote{For $\sigma_F=5\%$, there are two extra (the
  third and the fourth) eigenmodes with $c_\mu>0.01$ at $\ell=2000$
(Fig. \ref{fig:c}, open circle). But the resulting overestimation is
significantly larger than $(c_3+c_4)/(1-c_3-c_4)\simeq 0.05$ that we expect from the
$\lambda_\mu$-$c_\mu$ diagnostic. The reason is that now the
contamination to the lensing magnification bias has $N_F$ independent
components and the analytical result of
Eq. \ref{eqn:analyticalsolution} is no longer exact/unbiased.} Such systematic
error decreases with decreasing $\sigma_F$ and becomes negligible for
$\sigma_F\la 1\%$.   Therefore the ABS method may still work well
under this extreme condition. Nevertheless, it demonstrates that in lensing reconstruction
through cosmic magnification, flux calibration and its correction in
the galaxy number overdensity are important. Finally we remind that
the adopted model of extra systematic error in galaxy power spectra
measurement (Eq. \ref{eqn:flux}) is largely hypothetical and in reality such error may not
exist. The major purpose is to demonstrate the impact of extra
eigenmodes in the galaxy power spectra on the lensing reconstruction.

\section{Discussions and conclusions}
\label{sec:summary}
%%%%%%%%%%%%%%%%%%%%%%%%%%%%%%%%%%%%%%%%%%%%%%%%%%%%%%%%%%%%%%%%%%Tables
\begin{table}%[htb]
\caption{Cosmic magnification versus cosmic
  shear.  The analogy of galaxy intrinsic clustering in cosmic
  magnification measurement is galaxy intrinsic alignment in cosmic
  shear measurement.  Our ABS method can eliminate the galaxy
  intrinsic clustering, accurately and blindly.  }
\begin{tabular}{ccc} \\
Methods &cosmic shear& cosmic magnification \\ \hline\hline
observable & $\epsilon^L=\epsilon_g+\gamma$ &
$\delta_g^{L}=\delta_g+g\kappa$\\
signal  & shear $\gamma$ & convergence $\kappa$ \\
systematic error& intrinsic alignment & intrinsic clustering \\ 
&          (correlated $\epsilon_g)$ & (
correlated $\delta_g$)\\
statistical error & random shape noise & shot noise \\\hline\hline 
\end{tabular}
\label{table:comparison}
\end{table}

This paper outlines the ABS method to reconstruct weak lensing  power
spectrum by counting galaxies. It works on the measured cross galaxy power spectra between
different flux bins. Based on the analytical solution that we found,
the lensing power spectrum can be determined by a few straightforward linear
algebra operations. It does not rely on assumptions of galaxy
intrinsic clustering, making it robust against uncertainties in
modelling galaxy clustering. The only limiting factor is the galaxy
survey capability. To reliably determine the lensing signal, the
galaxy survey must be sufficiently wide and sufficiently
dense. Qualitatively speaking,  statistical fluctuations in the galaxy
power spectrum measurement (e.g. shot noise, but not cosmic variance)
must be subdominant to the lensing signal. This is 
challenging for stage III projects. But stage IV projects such as SKA
and LSST, with capability of measuring billion galaxies over half 
the sky, are promising to realize such lensing measurement
at $z\sim 1$ and $\ell\la 5\times 10^3$. Such measurement will be highly
complementary to cosmic shear measurement by the same surveys.

Table \ref{table:comparison} compares cosmic magnification and cosmic
shear. (1) Statistical error. Cosmic shear has the advantage of a factor of $3$ smaller statistical error per
galaxy.  But cosmic magnification has more galaxies to use, since
faint and small galaxies without useful cosmic shear measurement can
still be useful for cosmic magnification. Which has smaller statistical
error then depends on the survey specifications, galaxy luminosity
function and the threshold of selecting galaxies
(e.g. \citet{Zhang05}).  (2) Systematic error.  The 
analogy of galaxy intrinsic clustering in cosmic
magnification is the galaxy intrinsic alignment in cosmic
shear. There are intensive efforts to eliminate/alleviate the
intrinsic alignment (\citet{2015PhR...558....1T} and references 
therein).  One class of efforts is independent of intrinsic alignment
modelling,  such as nulling
\citep{2008A&A...488..829J,2009A&A...507..105J} and   self-calibration
\citep{Zhang10b,Zhang10c,2012MNRAS.419.1804T,2012MNRAS.423.1663T,2012MNRAS.427..442T}.
In contrast, there are significantly less efforts in dealing with the
intrinsic clustering.  This paper presents our latest result, the ABS
method, towards this
direction \citep{Zhang05,Zhang06a,YangXJ11,YangXJ15}. Similar to
self-calibration of intrinsic alignment, the ABS method does not rely on external data, nor
cosmological priors. It does not sacrifice cosmological information of
weak lensing, since it utilizes no information of  the lensing signal
$\kappa$. It only uses the information encoded in the prefactor $g$,
which by itself contains only astrophysical information of galaxy
luminosity function.  

Our ABS method is straightforward to implement in data 
analysis. What it does is to  post-process the observed galaxy power
spectra, whose measurement is a routine exercise.  Next we will target at 
specific surveys to make more realistic forecasts of the lensing reconstruction accuracy
by counting galaxies. We will combine with N-body simulations and
halo occupation distribution/conditional luminosity function
(e.g. \citet{Jing98,2003MNRAS.339.1057Y,2005ApJ...633..791Z}, to
generate more realistic $C^{\rm obs}_{ij}$.

%%%%%%%%%%%%%%%%%%%%%%%%%%%%%%%%%%%%%%%%%%%%%
\section*{Acknowledgments}
This work was supported by the National Science Foundation of China
(11603019, 11433001, 11621303, 11653003, 11320101002, 11403071, 11475148),  National
Basic Research Program of China (2015CB85701), Key Laboratory for Particle Physics,
Astrophysics and Cosmology, Ministry of Education, and Shanghai Key
Laboratory for Particle Physics and Cosmology (SKLPPC), and Zhejiang
province foundation for young researchers (LQ15A030001).

\bibliographystyle{apj}
\bibliography{mybib}

\begin{thebibliography}{}
\expandafter\ifx\csname natexlab\endcsname\relax\def\natexlab#1{#1}\fi

\bibitem[{{Ade} {et~al.}(2014{\natexlab{a}}){Ade}, {Akiba}, {Anthony},
  {Arnold}, {Atlas}, {Barron}, {Boettger}, {Borrill}, {Borys}, {Chapman},
  {Chinone}, {Dobbs}, {Elleflot}, {Errard}, {Fabbian}, {Feng}, {Flanigan},
  {Gilbert}, {Grainger}, {Halverson}, {Hasegawa}, {Hattori}, {Hazumi},
  {Holzapfel}, {Hori}, {Howard}, {Hyland}, {Inoue}, {Jaehnig}, {Jaffe},
  {Keating}, {Kermish}, {Keskitalo}, {Kisner}, {Le Jeune}, {Lee}, {Leitch},
  {Linder}, {Lungu}, {Matsuda}, {Matsumura}, {Meng}, {Miller}, {Morii},
  {Moyerman}, {Myers}, {Navaroli}, {Nishino}, {Paar}, {Peloton}, {Poletti},
  {Quealy}, {Rebeiz}, {Reichardt}, {Richards}, {Ross}, {Rotermund},
  {Schanning}, {Schenck}, {Sherwin}, {Shimizu}, {Shimmin}, {Shimon},
  {Siritanasak}, {Smecher}, {Spieler}, {Stebor}, {Steinbach}, {Stompor},
  {Suzuki}, {Takakura}, {Tikhomirov}, {Tomaru}, {Wilson}, {Yadav}, {Zahn}, \&
  {Polarbear Collaboration}}]{2014PhRvL.112m1302A}
{Ade}, P.~A.~R., {Akiba}, Y., {Anthony}, A.~E., {et~al.} 2014{\natexlab{a}},
  Physical Review Letters, 112, 131302

\bibitem[{{Ade} {et~al.}(2014{\natexlab{b}}){Ade}, {Akiba}, {Anthony},
  {Arnold}, {Atlas}, {Barron}, {Boettger}, {Borrill}, {Chapman}, {Chinone},
  {Dobbs}, {Elleflot}, {Errard}, {Fabbian}, {Feng}, {Flanigan}, {Gilbert},
  {Grainger}, {Halverson}, {Hasegawa}, {Hattori}, {Hazumi}, {Holzapfel},
  {Hori}, {Howard}, {Hyland}, {Inoue}, {Jaehnig}, {Jaffe}, {Keating},
  {Kermish}, {Keskitalo}, {Kisner}, {Le Jeune}, {Lee}, {Linder}, {Leitch},
  {Lungu}, {Matsuda}, {Matsumura}, {Meng}, {Miller}, {Morii}, {Moyerman},
  {Myers}, {Navaroli}, {Nishino}, {Paar}, {Peloton}, {Quealy}, {Rebeiz},
  {Reichardt}, {Richards}, {Ross}, {Schanning}, {Schenck}, {Sherwin},
  {Shimizu}, {Shimmin}, {Shimon}, {Siritanasak}, {Smecher}, {Spieler},
  {Stebor}, {Steinbach}, {Stompor}, {Suzuki}, {Takakura}, {Tomaru}, {Wilson},
  {Yadav}, {Zahn}, \& {Polarbear Collaboration}}]{2014PhRvL.113b1301A}
---. 2014{\natexlab{b}}, Physical Review Letters, 113, 021301

\bibitem[{{Albrecht} {et~al.}(2006){Albrecht}, {Bernstein}, {Cahn}, {Freedman},
  {Hewitt}, {Hu}, {Huth}, {Kamionkowski}, {Kolb}, {Knox}, {Mather}, {Staggs},
  \& {Suntzeff}}]{DETF}
{Albrecht}, A., {Bernstein}, G., {Cahn}, R., {et~al.} 2006, ArXiv Astrophysics
  e-prints, astro-ph/0609591

\bibitem[{{Alsing} {et~al.}(2015){Alsing}, {Kirk}, {Heavens}, \&
  {Jaffe}}]{2015MNRAS.452.1202A}
{Alsing}, J., {Kirk}, D., {Heavens}, A., \& {Jaffe}, A.~H. 2015, \mnras, 452,
  1202

\bibitem[{{Bartelmann}(1995)}]{1995A&A...298..661B}
{Bartelmann}, M. 1995, \aap, 298, 661

\bibitem[{{Bartelmann} \& {Schneider}(2001)}]{2001PhR...340..291B}
{Bartelmann}, M., \& {Schneider}, P. 2001, \physrep, 340, 291

\bibitem[{{Bauer} {et~al.}(2014){Bauer}, {Gazta{\~n}aga}, {Mart{\'{\i}}}, \&
  {Miquel}}]{2014MNRAS.440.3701B}
{Bauer}, A.~H., {Gazta{\~n}aga}, E., {Mart{\'{\i}}}, P., \& {Miquel}, R. 2014,
  \mnras, 440, 3701

\bibitem[{{Becker} {et~al.}(2016){Becker}, {Troxel}, {MacCrann}, {Krause},
  {Eifler}, {Friedrich}, {Nicola}, {Refregier}, {Amara}, {Bacon}, {Bernstein},
  {Bonnett}, {Bridle}, {Busha}, {Chang}, {Dodelson}, {Erickson}, {Evrard},
  {Frieman}, {Gaztanaga}, {Gruen}, {Hartley}, {Jain}, {Jarvis}, {Kacprzak},
  {Kirk}, {Kravtsov}, {Leistedt}, {Peiris}, {Rykoff}, {Sabiu}, {S{\'a}nchez},
  {Seo}, {Sheldon}, {Wechsler}, {Zuntz}, {Abbott}, {Abdalla}, {Allam},
  {Armstrong}, {Banerji}, {Bauer}, {Benoit-L{\'e}vy}, {Bertin}, {Brooks},
  {Buckley-Geer}, {Burke}, {Capozzi}, {Carnero Rosell}, {Carrasco Kind},
  {Carretero}, {Castander}, {Crocce}, {Cunha}, {D'Andrea}, {da Costa}, {DePoy},
  {Desai}, {Diehl}, {Dietrich}, {Doel}, {Fausti Neto}, {Fernandez}, {Finley},
  {Flaugher}, {Fosalba}, {Gerdes}, {Gruendl}, {Gutierrez}, {Honscheid},
  {James}, {Kuehn}, {Kuropatkin}, {Lahav}, {Li}, {Lima}, {Maia}, {March},
  {Martini}, {Melchior}, {Miller}, {Miquel}, {Mohr}, {Nichol}, {Nord},
  {Ogando}, {Plazas}, {Reil}, {Romer}, {Roodman}, {Sako}, {Sanchez},
  {Scarpine}, {Schubnell}, {Sevilla-Noarbe}, {Smith}, {Soares-Santos},
  {Sobreira}, {Suchyta}, {Swanson}, {Tarle}, {Thaler}, {Thomas}, {Vikram},
  {Walker}, \& {Dark Energy Survey Collaboration}}]{2016PhRvD..94b2002B}
{Becker}, M.~R., {Troxel}, M.~A., {MacCrann}, N., {et~al.} 2016, \prd, 94,
  022002

\bibitem[{{Bonoli} \& {Pen}(2009)}]{Bonoli09}
{Bonoli}, S., \& {Pen}, U.~L. 2009, \mnras, 396, 1610

\bibitem[{{Chang} {et~al.}(2015){Chang}, {Vikram}, {Jain}, {Bacon}, {Amara},
  {Becker}, {Bernstein}, {Bonnett}, {Bridle}, {Brout}, {Busha}, {Frieman},
  {Gaztanaga}, {Hartley}, {Jarvis}, {Kacprzak}, {Kov{\'a}cs}, {Lahav}, {Lin},
  {Melchior}, {Peiris}, {Rozo}, {Rykoff}, {S{\'a}nchez}, {Sheldon}, {Troxel},
  {Wechsler}, {Zuntz}, {Abbott}, {Abdalla}, {Allam}, {Annis}, {Bauer},
  {Benoit-L{\'e}vy}, {Brooks}, {Buckley-Geer}, {Burke}, {Capozzi}, {Carnero
  Rosell}, {Carrasco Kind}, {Castander}, {Crocce}, {D'Andrea}, {Desai},
  {Diehl}, {Dietrich}, {Doel}, {Eifler}, {Evrard}, {Fausti Neto}, {Flaugher},
  {Fosalba}, {Gruen}, {Gruendl}, {Gutierrez}, {Honscheid}, {James}, {Kent},
  {Kuehn}, {Kuropatkin}, {Maia}, {March}, {Martini}, {Merritt}, {Miller},
  {Miquel}, {Neilsen}, {Nichol}, {Ogando}, {Plazas}, {Romer}, {Roodman},
  {Sako}, {Sanchez}, {Sevilla}, {Smith}, {Soares-Santos}, {Sobreira},
  {Suchyta}, {Tarle}, {Thaler}, {Thomas}, {Tucker}, \&
  {Walker}}]{2015PhRvL.115e1301C}
{Chang}, C., {Vikram}, V., {Jain}, B., {et~al.} 2015, Physical Review Letters,
  115, 051301

\bibitem[{{Chiu} {et~al.}(2016){Chiu}, {Dietrich}, {Mohr}, {Applegate},
  {Benson}, {Bleem}, {Bayliss}, {Bocquet}, {Carlstrom}, {Capasso}, {Desai},
  {Gangkofner}, {Gonzalez}, {Gupta}, {Hennig}, {Hoekstra}, {von der Linden},
  {Liu}, {McDonald}, {Reichardt}, {Saro}, {Schrabback}, {Strazzullo}, {Stubbs},
  \& {Zenteno}}]{2016MNRAS.457.3050C}
{Chiu}, I., {Dietrich}, J.~P., {Mohr}, J., {et~al.} 2016, \mnras, 457, 3050

\bibitem[{{Cooray} {et~al.}(2006){Cooray}, {Holz}, \& {Huterer}}]{Cooray06a}
{Cooray}, A., {Holz}, D.~E., \& {Huterer}, D. 2006, \apjl, 637, L77

\bibitem[{{Das} {et~al.}(2011){Das}, {Sherwin}, {Aguirre}, {Appel}, {Bond},
  {Carvalho}, {Devlin}, {Dunkley}, {D{\"u}nner}, {Essinger-Hileman}, {Fowler},
  {Hajian}, {Halpern}, {Hasselfield}, {Hincks}, {Hlozek}, {Huffenberger},
  {Hughes}, {Irwin}, {Klein}, {Kosowsky}, {Lupton}, {Marriage}, {Marsden},
  {Menanteau}, {Moodley}, {Niemack}, {Nolta}, {Page}, {Parker}, {Reese},
  {Schmitt}, {Sehgal}, {Sievers}, {Spergel}, {Staggs}, {Swetz}, {Switzer},
  {Thornton}, {Visnjic}, \& {Wollack}}]{2011PhRvL.107b1301D}
{Das}, S., {Sherwin}, B.~D., {Aguirre}, P., {et~al.} 2011, Physical Review
  Letters, 107, 021301

\bibitem[{{Dodelson} \& {Vallinotto}(2006)}]{2006PhRvD..74f3515D}
{Dodelson}, S., \& {Vallinotto}, A. 2006, \prd, 74, 063515

\bibitem[{{Duncan} {et~al.}(2016){Duncan}, {Heymans}, {Heavens}, \&
  {Joachimi}}]{2016MNRAS.457..764D}
{Duncan}, C.~A.~J., {Heymans}, C., {Heavens}, A.~F., \& {Joachimi}, B. 2016,
  \mnras, 457, 764

\bibitem[{{Duncan} {et~al.}(2014){Duncan}, {Joachimi}, {Heavens}, {Heymans}, \&
  {Hildebrandt}}]{2014MNRAS.437.2471D}
{Duncan}, C.~A.~J., {Joachimi}, B., {Heavens}, A.~F., {Heymans}, C., \&
  {Hildebrandt}, H. 2014, \mnras, 437, 2471

\bibitem[{{Ford} {et~al.}(2014){Ford}, {Hildebrandt}, {Van Waerbeke}, {Erben},
  {Laigle}, {Milkeraitis}, \& {Morrison}}]{2014MNRAS.439.3755F}
{Ford}, J., {Hildebrandt}, H., {Van Waerbeke}, L., {et~al.} 2014, \mnras, 439,
  3755

\bibitem[{{Fu} {et~al.}(2014){Fu}, {Kilbinger}, {Erben}, {Heymans},
  {Hildebrandt}, {Hoekstra}, {Kitching}, {Mellier}, {Miller}, {Semboloni},
  {Simon}, {Van Waerbeke}, {Coupon}, {Harnois-D{\'e}raps}, {Hudson}, {Kuijken},
  {Rowe}, {Schrabback}, {Vafaei}, \& {Velander}}]{2014MNRAS.441.2725F}
{Fu}, L., {Kilbinger}, M., {Erben}, T., {et~al.} 2014, \mnras, 441, 2725

\bibitem[{{Garcia-Fernandez} {et~al.}(2016){Garcia-Fernandez}, {S{\'a}nchez},
  {Sevilla-Noarbe}, {Suchyta}, {Huff}, {Gaztanaga}, {Aleksi{\'c}}, {Ponce},
  {Castander}, {Hoyle}, {Abbott}, {Abdalla}, {Allam}, {Annis},
  {Benoit-L{\'e}vy}, {Bernstein}, {Bertin}, {Brooks}, {Buckley-Geer}, {Burke},
  {Carnero Rosell}, {Carrasco Kind}, {Carretero}, {Crocce}, {Cunha},
  {D'Andrea}, {da Costa}, {DePoy}, {Desai}, {Diehl}, {Eifler}, {Evrard},
  {Fernandez}, {Flaugher}, {Fosalba}, {Frieman}, {Garc{\'{\i}}a-Bellido},
  {Gerdes}, {Giannantonio}, {Gruen}, {Gruendl}, {Gschwend}, {Gutierrez},
  {James}, {Jarvis}, {Kirk}, {Krause}, {Kuehn}, {Kuropatkin}, {Lahav}, {Lima},
  {MacCrann}, {Maia}, {March}, {Marshall}, {Melchior}, {Miquel}, {Mohr},
  {Plazas}, {Romer}, {Roodman}, {Rykoff}, {Scarpine}, {Schubnell}, {Smith},
  {Soares-Santos}, {Sobreira}, {Tarle}, {Thomas}, {Walker}, \&
  {Wester}}]{2016arXiv161110326G}
{Garcia-Fernandez}, M., {S{\'a}nchez}, E., {Sevilla-Noarbe}, I., {et~al.} 2016,
  ArXiv e-prints, arXiv:1611.10326

\bibitem[{{Gonz{\'a}lez-Nuevo} {et~al.}(2014){Gonz{\'a}lez-Nuevo}, {Lapi},
  {Negrello}, {Danese}, {De Zotti}, {Amber}, {Baes}, {Bland-Hawthorn},
  {Bourne}, {Brough}, {Bussmann}, {Cai}, {Cooray}, {Driver}, {Dunne}, {Dye},
  {Eales}, {Ibar}, {Ivison}, {Liske}, {Loveday}, {Maddox}, {Micha{\l}owski},
  {Robotham}, {Scott}, {Smith}, {Valiante}, \& {Xia}}]{2014MNRAS.442.2680G}
{Gonz{\'a}lez-Nuevo}, J., {Lapi}, A., {Negrello}, M., {et~al.} 2014, \mnras,
  442, 2680

\bibitem[{{Hamaus} {et~al.}(2010){Hamaus}, {Seljak}, {Desjacques}, {Smith}, \&
  {Baldauf}}]{Hamaus10}
{Hamaus}, N., {Seljak}, U., {Desjacques}, V., {Smith}, R.~E., \& {Baldauf}, T.
  2010, \prd, 82, 043515

\bibitem[{{Hanson} {et~al.}(2013){Hanson}, {Hoover}, {Crites}, {Ade}, {Aird},
  {Austermann}, {Beall}, {Bender}, {Benson}, {Bleem}, {Bock}, {Carlstrom},
  {Chang}, {Chiang}, {Cho}, {Conley}, {Crawford}, {de Haan}, {Dobbs},
  {Everett}, {Gallicchio}, {Gao}, {George}, {Halverson}, {Harrington},
  {Henning}, {Hilton}, {Holder}, {Holzapfel}, {Hrubes}, {Huang}, {Hubmayr},
  {Irwin}, {Keisler}, {Knox}, {Lee}, {Leitch}, {Li}, {Liang}, {Luong-Van},
  {Marsden}, {McMahon}, {Mehl}, {Meyer}, {Mocanu}, {Montroy}, {Natoli},
  {Nibarger}, {Novosad}, {Padin}, {Pryke}, {Reichardt}, {Ruhl}, {Saliwanchik},
  {Sayre}, {Schaffer}, {Schulz}, {Smecher}, {Stark}, {Story}, {Tucker},
  {Vanderlinde}, {Vieira}, {Viero}, {Wang}, {Yefremenko}, {Zahn}, \&
  {Zemcov}}]{2013PhRvL.111n1301H}
{Hanson}, D., {Hoover}, S., {Crites}, A., {et~al.} 2013, Physical Review
  Letters, 111, 141301

\bibitem[{{Heavens} {et~al.}(2013){Heavens}, {Alsing}, \&
  {Jaffe}}]{2013MNRAS.433L...6H}
{Heavens}, A., {Alsing}, J., \& {Jaffe}, A.~H. 2013, \mnras, 433, L6

\bibitem[{{Heymans} {et~al.}(2012){Heymans}, {Van Waerbeke}, {Miller}, {Erben},
  {Hildebrandt}, {Hoekstra}, {Kitching}, {Mellier}, {Simon}, {Bonnett},
  {Coupon}, {Fu}, {Harnois D{\'e}raps}, {Hudson}, {Kilbinger}, {Kuijken},
  {Rowe}, {Schrabback}, {Semboloni}, {van Uitert}, {Vafaei}, \&
  {Velander}}]{2012MNRAS.427..146H}
{Heymans}, C., {Van Waerbeke}, L., {Miller}, L., {et~al.} 2012, \mnras, 427,
  146

\bibitem[{{Hildebrandt} {et~al.}(2009){Hildebrandt}, {van Waerbeke}, \&
  {Erben}}]{2009A&A...507..683H}
{Hildebrandt}, H., {van Waerbeke}, L., \& {Erben}, T. 2009, \aap, 507, 683

\bibitem[{{Hildebrandt} {et~al.}(2013){Hildebrandt}, {van Waerbeke}, {Scott},
  {B{\'e}thermin}, {Bock}, {Clements}, {Conley}, {Cooray}, {Dunlop}, {Eales},
  {Erben}, {Farrah}, {Franceschini}, {Glenn}, {Halpern}, {Heinis}, {Ivison},
  {Marsden}, {Oliver}, {Page}, {P{\'e}rez-Fournon}, {Smith}, {Rowan-Robinson},
  {Valtchanov}, {van der Burg}, {Vieira}, {Viero}, \&
  {Wang}}]{2013MNRAS.429.3230H}
{Hildebrandt}, H., {van Waerbeke}, L., {Scott}, D., {et~al.} 2013, \mnras, 429,
  3230

\bibitem[{{Hildebrandt} {et~al.}(2016){Hildebrandt}, {Choi}, {Heymans},
  {Blake}, {Erben}, {Miller}, {Nakajima}, {van Waerbeke}, {Viola},
  {Buddendiek}, {Harnois-D{\'e}raps}, {Hojjati}, {Joachimi}, {Joudaki},
  {Kitching}, {Wolf}, {Gwyn}, {Johnson}, {Kuijken}, {Sheikhbahaee}, {Tudorica},
  \& {Yee}}]{2016MNRAS.463..635H}
{Hildebrandt}, H., {Choi}, A., {Heymans}, C., {et~al.} 2016, \mnras, 463, 635

\bibitem[{{Hildebrandt} {et~al.}(2017){Hildebrandt}, {Viola}, {Heymans},
  {Joudaki}, {Kuijken}, {Blake}, {Erben}, {Joachimi}, {Klaes}, {Miller},
  {Morrison}, {Nakajima}, {Verdoes Kleijn}, {Amon}, {Choi}, {Covone}, {de
  Jong}, {Dvornik}, {Fenech Conti}, {Grado}, {Harnois-D{\'e}raps}, {Herbonnet},
  {Hoekstra}, {K{\"o}hlinger}, {McFarland}, {Mead}, {Merten}, {Napolitano},
  {Peacock}, {Radovich}, {Schneider}, {Simon}, {Valentijn}, {van den Busch},
  {van Uitert}, \& {Van Waerbeke}}]{2017MNRAS.465.1454H}
{Hildebrandt}, H., {Viola}, M., {Heymans}, C., {et~al.} 2017, \mnras, 465, 1454

\bibitem[{{Hirata} {et~al.}(2008){Hirata}, {Ho}, {Padmanabhan}, {Seljak}, \&
  {Bahcall}}]{2008PhRvD..78d3520H}
{Hirata}, C.~M., {Ho}, S., {Padmanabhan}, N., {Seljak}, U., \& {Bahcall}, N.~A.
  2008, \prd, 78, 043520

\bibitem[{{Hoekstra} \& {Jain}(2008)}]{Hoekstra08}
{Hoekstra}, H., \& {Jain}, B. 2008, Annual Review of Nuclear and Particle
  Science, 58, 99

\bibitem[{{Huff} {et~al.}(2014){Huff}, {Eifler}, {Hirata}, {Mandelbaum},
  {Schlegel}, \& {Seljak}}]{2014MNRAS.440.1322H}
{Huff}, E.~M., {Eifler}, T., {Hirata}, C.~M., {et~al.} 2014, \mnras, 440, 1322

\bibitem[{{Huff} \& {Graves}(2014)}]{2014ApJ...780L..16H}
{Huff}, E.~M., \& {Graves}, G.~J. 2014, \apjl, 780, L16

\bibitem[{{Jarvis} {et~al.}(2016){Jarvis}, {Sheldon}, {Zuntz}, {Kacprzak},
  {Bridle}, {Amara}, {Armstrong}, {Becker}, {Bernstein}, {Bonnett}, {Chang},
  {Das}, {Dietrich}, {Drlica-Wagner}, {Eifler}, {Gangkofner}, {Gruen},
  {Hirsch}, {Huff}, {Jain}, {Kent}, {Kirk}, {MacCrann}, {Melchior}, {Plazas},
  {Refregier}, {Rowe}, {Rykoff}, {Samuroff}, {S{\'a}nchez}, {Suchyta},
  {Troxel}, {Vikram}, {Abbott}, {Abdalla}, {Allam}, {Annis}, {Benoit-L{\'e}vy},
  {Bertin}, {Brooks}, {Buckley-Geer}, {Burke}, {Capozzi}, {Carnero Rosell},
  {Carrasco Kind}, {Carretero}, {Castander}, {Clampitt}, {Crocce}, {Cunha},
  {D'Andrea}, {da Costa}, {DePoy}, {Desai}, {Diehl}, {Doel}, {Fausti Neto},
  {Flaugher}, {Fosalba}, {Frieman}, {Gaztanaga}, {Gerdes}, {Gruendl},
  {Gutierrez}, {Honscheid}, {James}, {Kuehn}, {Kuropatkin}, {Lahav}, {Li},
  {Lima}, {March}, {Martini}, {Miquel}, {Mohr}, {Neilsen}, {Nord}, {Ogando},
  {Reil}, {Romer}, {Roodman}, {Sako}, {Sanchez}, {Scarpine}, {Schubnell},
  {Sevilla-Noarbe}, {Smith}, {Soares-Santos}, {Sobreira}, {Swanson}, {Tarle},
  {Thaler}, {Thomas}, {Walker}, \& {Wechsler}}]{2016MNRAS.460.2245J}
{Jarvis}, M., {Sheldon}, E., {Zuntz}, J., {et~al.} 2016, \mnras, 460, 2245

\bibitem[{{Jing} {et~al.}(1998){Jing}, {Mo}, \& {Boerner}}]{Jing98}
{Jing}, Y.~P., {Mo}, H.~J., \& {Boerner}, G. 1998, \apj, 494, 1

\bibitem[{{Jing} {et~al.}(2006){Jing}, {Zhang}, {Lin}, {Gao}, \&
  {Springel}}]{2006ApJ...640L.119J}
{Jing}, Y.~P., {Zhang}, P., {Lin}, W.~P., {Gao}, L., \& {Springel}, V. 2006,
  \apjl, 640, L119

\bibitem[{{Joachimi} \& {Schneider}(2008)}]{2008A&A...488..829J}
{Joachimi}, B., \& {Schneider}, P. 2008, \aap, 488, 829

\bibitem[{{Joachimi} \& {Schneider}(2009)}]{2009A&A...507..105J}
---. 2009, \aap, 507, 105

\bibitem[{Kilbinger {et~al.}(2013)}]{Kilbinger:2012qz}
Kilbinger, M., {et~al.} 2013, Mon. Not. Roy. Astron. Soc., 430, 2200

\bibitem[{{Mandelbaum} {et~al.}(2015){Mandelbaum}, {Rowe}, {Armstrong}, {Bard},
  {Bertin}, {Bosch}, {Boutigny}, {Courbin}, {Dawson}, {Donnarumma}, {Fenech
  Conti}, {Gavazzi}, {Gentile}, {Gill}, {Hogg}, {Huff}, {Jee}, {Kacprzak},
  {Kilbinger}, {Kuntzer}, {Lang}, {Luo}, {March}, {Marshall}, {Meyers},
  {Miller}, {Miyatake}, {Nakajima}, {Ngol{\'e} Mboula}, {Nurbaeva}, {Okura},
  {Paulin-Henriksson}, {Rhodes}, {Schneider}, {Shan}, {Sheldon}, {Simet},
  {Starck}, {Sureau}, {Tewes}, {Zarb Adami}, {Zhang}, \&
  {Zuntz}}]{2015MNRAS.450.2963M}
{Mandelbaum}, R., {Rowe}, B., {Armstrong}, R., {et~al.} 2015, \mnras, 450, 2963

\bibitem[{{Matsubara}(2008)}]{Matsubara08a}
{Matsubara}, T. 2008, \prd, 77, 063530

\bibitem[{{M{\'e}nard} {et~al.}(2010){M{\'e}nard}, {Scranton}, {Fukugita}, \&
  {Richards}}]{2010MNRAS.405.1025M}
{M{\'e}nard}, B., {Scranton}, R., {Fukugita}, M., \& {Richards}, G. 2010,
  \mnras, 405, 1025

\bibitem[{{Morrison} {et~al.}(2012){Morrison}, {Scranton}, {M{\'e}nard},
  {Schmidt}, {Tyson}, {Ryan}, {Choi}, \& {Wittman}}]{2012MNRAS.426.2489M}
{Morrison}, C.~B., {Scranton}, R., {M{\'e}nard}, B., {et~al.} 2012, \mnras,
  426, 2489

\bibitem[{{Munshi} {et~al.}(2008){Munshi}, {Valageas}, {van Waerbeke}, \&
  {Heavens}}]{Munshi08}
{Munshi}, D., {Valageas}, P., {van Waerbeke}, L., \& {Heavens}, A. 2008,
  \physrep, 462, 67

\bibitem[{{Nishizawa} {et~al.}(2013){Nishizawa}, {Takada}, \&
  {Nishimichi}}]{Nishizawa13}
{Nishizawa}, A.~J., {Takada}, M., \& {Nishimichi}, T. 2013, \mnras, 433, 209

\bibitem[{{Okamoto} \& {Hu}(2003)}]{2003PhRvD..67h3002O}
{Okamoto}, T., \& {Hu}, W. 2003, \prd, 67, 083002

\bibitem[{{Padmanabhan} \& {White}(2009)}]{Padmanabhan09}
{Padmanabhan}, N., \& {White}, M. 2009, \prd, 80, 063508

\bibitem[{{Planck Collaboration} {et~al.}(2014){Planck Collaboration}, {Ade},
  {Aghanim}, {Armitage-Caplan}, {Arnaud}, {Ashdown}, {Atrio-Barandela},
  {Aumont}, {Baccigalupi}, {Banday}, \& et~al.}]{2014A&A...571A..17P}
{Planck Collaboration}, {Ade}, P.~A.~R., {Aghanim}, N., {et~al.} 2014, \aap,
  571, A17

\bibitem[{{Planck Collaboration} {et~al.}(2016){Planck Collaboration}, {Ade},
  {Aghanim}, {Arnaud}, {Ashdown}, {Aumont}, {Baccigalupi}, {Banday},
  {Barreiro}, {Bartlett}, \& et~al.}]{2016A&A...594A..15P}
---. 2016, \aap, 594, A15

\bibitem[{{Refregier}(2003)}]{2003ARA&A..41..645R}
{Refregier}, A. 2003, \araa, 41, 645

\bibitem[{{Rudd} {et~al.}(2008){Rudd}, {Zentner}, \&
  {Kravtsov}}]{2008ApJ...672...19R}
{Rudd}, D.~H., {Zentner}, A.~R., \& {Kravtsov}, A.~V. 2008, \apj, 672, 19

\bibitem[{{Schmidt} {et~al.}(2012){Schmidt}, {Leauthaud}, {Massey}, {Rhodes},
  {George}, {Koekemoer}, {Finoguenov}, \& {Tanaka}}]{2012ApJ...744L..22S}
{Schmidt}, F., {Leauthaud}, A., {Massey}, R., {et~al.} 2012, \apjl, 744, L22

\bibitem[{{Scranton} {et~al.}(2005){Scranton}, {M{\'e}nard}, {Richards},
  {Nichol}, {Myers}, {Jain}, {Gray}, {Bartelmann}, {Brunner}, {Connolly},
  {Gunn}, {Sheth}, {Bahcall}, {Brinkman}, {Loveday}, {Schneider}, {Thakar}, \&
  {York}}]{2005ApJ...633..589S}
{Scranton}, R., {M{\'e}nard}, B., {Richards}, G.~T., {et~al.} 2005, \apj, 633,
  589

\bibitem[{{Seljak}(1996)}]{Seljak96}
{Seljak}, U. 1996, \apj, 463, 1

\bibitem[{{Seljak} \& {Zaldarriaga}(1999)}]{1999PhRvL..82.2636S}
{Seljak}, U., \& {Zaldarriaga}, M. 1999, Physical Review Letters, 82, 2636

\bibitem[{{Smith} {et~al.}(2007){Smith}, {Zahn}, \&
  {Dor{\'e}}}]{2007PhRvD..76d3510S}
{Smith}, K.~M., {Zahn}, O., \& {Dor{\'e}}, O. 2007, \prd, 76, 043510

\bibitem[{{Smith} {et~al.}(2003){Smith}, {Peacock}, {Jenkins}, {White},
  {Frenk}, {Pearce}, {Thomas}, {Efstathiou}, \& {Couchman}}]{Smith03}
{Smith}, R.~E., {Peacock}, J.~A., {Jenkins}, A., {et~al.} 2003, \mnras, 341,
  1311

\bibitem[{{Troxel} \& {Ishak}(2012{\natexlab{a}})}]{2012MNRAS.427..442T}
{Troxel}, M.~A., \& {Ishak}, M. 2012{\natexlab{a}}, \mnras, 427, 442

\bibitem[{{Troxel} \& {Ishak}(2012{\natexlab{b}})}]{2012MNRAS.423.1663T}
---. 2012{\natexlab{b}}, \mnras, 423, 1663

\bibitem[{{Troxel} \& {Ishak}(2012{\natexlab{c}})}]{2012MNRAS.419.1804T}
---. 2012{\natexlab{c}}, \mnras, 419, 1804

\bibitem[{{Troxel} \& {Ishak}(2015)}]{2015PhR...558....1T}
---. 2015, \physrep, 558, 1

\bibitem[{{Vallinotto} {et~al.}(2011){Vallinotto}, {Dodelson}, \&
  {Zhang}}]{Vallinotto11}
{Vallinotto}, A., {Dodelson}, S., \& {Zhang}, P. 2011, \prd, 84, 103004

\bibitem[{{van Engelen} {et~al.}(2012){van Engelen}, {Keisler}, {Zahn}, {Aird},
  {Benson}, {Bleem}, {Carlstrom}, {Chang}, {Cho}, {Crawford}, {Crites}, {de
  Haan}, {Dobbs}, {Dudley}, {George}, {Halverson}, {Holder}, {Holzapfel},
  {Hoover}, {Hou}, {Hrubes}, {Joy}, {Knox}, {Lee}, {Leitch}, {Lueker},
  {Luong-Van}, {McMahon}, {Mehl}, {Meyer}, {Millea}, {Mohr}, {Montroy},
  {Natoli}, {Padin}, {Plagge}, {Pryke}, {Reichardt}, {Ruhl}, {Sayre},
  {Schaffer}, {Shaw}, {Shirokoff}, {Spieler}, {Staniszewski}, {Stark}, {Story},
  {Vanderlinde}, {Vieira}, \& {Williamson}}]{2012ApJ...756..142V}
{van Engelen}, A., {Keisler}, R., {Zahn}, O., {et~al.} 2012, \apj, 756, 142

\bibitem[{{van Engelen} {et~al.}(2015){van Engelen}, {Sherwin}, {Sehgal},
  {Addison}, {Allison}, {Battaglia}, {de Bernardis}, {Bond}, {Calabrese},
  {Coughlin}, {Crichton}, {Datta}, {Devlin}, {Dunkley}, {D{\"u}nner},
  {Gallardo}, {Grace}, {Gralla}, {Hajian}, {Hasselfield}, {Henderson}, {Hill},
  {Hilton}, {Hincks}, {Hlozek}, {Huffenberger}, {Hughes}, {Koopman},
  {Kosowsky}, {Louis}, {Lungu}, {Madhavacheril}, {Maurin}, {McMahon},
  {Moodley}, {Munson}, {Naess}, {Nati}, {Newburgh}, {Niemack}, {Nolta}, {Page},
  {Pappas}, {Partridge}, {Schmitt}, {Sievers}, {Simon}, {Spergel}, {Staggs},
  {Switzer}, {Ward}, \& {Wollack}}]{2015ApJ...808....7V}
{van Engelen}, A., {Sherwin}, B.~D., {Sehgal}, N., {et~al.} 2015, \apj, 808, 7

\bibitem[{{Wang} {et~al.}(2011){Wang}, {Cooray}, {Farrah}, {Amblard}, {Auld},
  {Bock}, {Brisbin}, {Burgarella}, {Chanial}, {Clements}, {Eales},
  {Franceschini}, {Glenn}, {Gong}, {Griffin}, {Heinis}, {Ibar}, {Ivison},
  {Mortier}, {Oliver}, {Page}, {Papageorgiou}, {Pearson}, {P{\'e}rez-Fournon},
  {Pohlen}, {Rawlings}, {Raymond}, {Rodighiero}, {Roseboom}, {Rowan-Robinson},
  {Scott}, {Serra}, {Seymour}, {Smith}, {Symeonidis}, {Tugwell}, {Vaccari},
  {Vieira}, {Vigroux}, \& {Wright}}]{2011MNRAS.414..596W}
{Wang}, L., {Cooray}, A., {Farrah}, D., {et~al.} 2011, \mnras, 414, 596

\bibitem[{{Weinberg} {et~al.}(2013){Weinberg}, {Mortonson}, {Eisenstein},
  {Hirata}, {Riess}, \& {Rozo}}]{2013PhR...530...87W}
{Weinberg}, D.~H., {Mortonson}, M.~J., {Eisenstein}, D.~J., {et~al.} 2013,
  \physrep, 530, 87

\bibitem[{{White}(2004)}]{2004APh....22..211W}
{White}, M. 2004, Astroparticle Physics, 22, 211

\bibitem[{{Wyithe} {et~al.}(2011){Wyithe}, {Yan}, {Windhorst}, \&
  {Mao}}]{2011Natur.469..181W}
{Wyithe}, J.~S.~B., {Yan}, H., {Windhorst}, R.~A., \& {Mao}, S. 2011, \nat,
  469, 181

\bibitem[{{Yang} {et~al.}(2003){Yang}, {Mo}, \& {van den
  Bosch}}]{2003MNRAS.339.1057Y}
{Yang}, X., {Mo}, H.~J., \& {van den Bosch}, F.~C. 2003, \mnras, 339, 1057

\bibitem[{{Yang} \& {Zhang}(2011)}]{YangXJ11}
{Yang}, X., \& {Zhang}, P. 2011, \mnras, 415, 3485

\bibitem[{{Yang} {et~al.}(2015){Yang}, {Zhang}, {Zhang}, \& {Yu}}]{YangXJ15}
{Yang}, X., {Zhang}, P., {Zhang}, J., \& {Yu}, Y. 2015, \mnras, 447, 345

\bibitem[{{Zhan} \& {Knox}(2004)}]{2004ApJ...616L..75Z}
{Zhan}, H., \& {Knox}, L. 2004, \apjl, 616, L75

\bibitem[{{Zhang}(2010{\natexlab{a}})}]{Zhang10b}
{Zhang}, P. 2010{\natexlab{a}}, \mnras, 406, L95

\bibitem[{{Zhang}(2010{\natexlab{b}})}]{Zhang10c}
---. 2010{\natexlab{b}}, \apj, 720, 1090

\bibitem[{{Zhang}(2015)}]{2015ApJ...806...45Z}
---. 2015, \apj, 806, 45

\bibitem[{{Zhang} \& {Pen}(2005)}]{Zhang05}
{Zhang}, P., \& {Pen}, U.-L. 2005, Physical Review Letters, 95, 241302

\bibitem[{{Zhang} \& {Pen}(2006)}]{Zhang06a}
---. 2006, \mnras, 367, 169

\bibitem[{{Zhang} {et~al.}(2016){Zhang}, {Zhang}, \& {Zhang}}]{ABS}
{Zhang}, P., {Zhang}, J., \& {Zhang}, L. 2016, ArXiv e-prints, arXiv:1608.03707

\bibitem[{{Zheng} {et~al.}(2005){Zheng}, {Berlind}, {Weinberg}, {Benson},
  {Baugh}, {Cole}, {Dav{\'e}}, {Frenk}, {Katz}, \&
  {Lacey}}]{2005ApJ...633..791Z}
{Zheng}, Z., {Berlind}, A.~A., {Weinberg}, D.~H., {et~al.} 2005, \apj, 633, 791

\end{thebibliography}
\appendix
\section{The fiducial model of galaxy biases}
\label{sec:bias}
The two biases ($b^{(1,2)}$) are given in term of the derivations of the halo mass
function \citep{Matsubara08a, Padmanabhan09, Nishizawa13}, \ba
b^{(1)}(M_{\rm
DM},z)&=&\frac{1}{\delta_c}\left[\nu^2-1+\frac{2p}{1+(q\nu^2)^p}\right]+1\
, \\ \nonumber b^{(2)}(M_{\rm
DM},z)&=&\frac{1}{\delta_c^2}\left[q^2\nu^4-3q\nu^2+\frac{2p(2q\nu^2+2p-1)}{1+(q\nu^2)^p}\right]+\frac{8}{21}(b^{(1)}-1)\ . \ea
Where $\nu=\delta_c(z)/\sigma(M_{\rm DM})$. $\sigma(M_{\rm DM})$ is
the linearly evolved $\rm rms$ density fluctuation with a top-hat
window function and $\delta_c(z)=1.686/D(z)$. Here, $p$ and $q$ are
parameters in the mass functions. The PS mass function is given with
$p=0$ and $q=1$. HI galaxies are selected by their neutral hydrogen
mass $M_{\rm HI}$. To calculate the associated biases of these
galaxies, we need to convert $M_{\rm HI}$ to $M_{\rm DM}$. We simply
choose $f_{\rm HI}=M_{\rm HI}/M_{\rm DM}=0.1$.  We caution that the adopted bias
model has several simplifications, and therefore may not well
represent that in realistic survey. First, we have neglected the
$\delta_m$-$\delta_m^2$ cross correlation, which vanishes at large
scales. Second, in principle we should also include the cubic bias
since it contributes comparably to the power spectrum as the
quadratic bias. Third,  $f_{\rm HI}$ should depend on the halo
mass. In the main text we have extended the above bias model to much
general cases, and shown that our ABS method of lensing reconstruction
is insensitive to details of galaxy bias. 

\end{document}